\documentclass[twocolumn,showpacs,amsmath,amssymb,aps,pra,floatfix,nofootinbib,superscriptaddress]{revtex4}
\usepackage{graphicx}
\usepackage{braket}
\usepackage{amsmath}
\DeclareMathOperator{\Tr}{Tr}

\begin{document}
\title{Unravelling Quantum Dot Array Simulators via Singlet-Triplet Measurements}
\author{Johnnie Gray}
\affiliation{Department of Physics and Astronomy, University College London,
Gower Street, London WC1E 6BT, United Kingdom}
\email{john.gray.14@ucl.ac.uk}
\author{Abolfazl Bayat}
\affiliation{Department of Physics and Astronomy, University College London,
	Gower Street, London WC1E 6BT, United Kingdom}
\author{Reuben K. Puddy}
\affiliation{Cavendish Laboratory, University of Cambridge, Cambridge CB3 0HE, United Kingdom}
\author{Charles G. Smith}
\affiliation{Cavendish Laboratory, University of Cambridge, Cambridge CB3 0HE, United Kingdom}
\author{Sougato Bose}
\affiliation{Department of Physics and Astronomy, University College London,
	Gower Street, London WC1E 6BT, United Kingdom}
\date{\today}

\begin{abstract}
Recently, singlet-triplet measurements in double dots have emerged as a powerful tool in quantum information processing.
In parallel, quantum dot arrays are being envisaged as analog quantum simulators of many-body models.
Thus motivated, we explore the potential of the above singlet-triplet measurements for probing and exploiting the ground-state of a Heisenberg spin chain in such a quantum simulator.
We formulate an efficient protocol to discriminate the achieved many-body ground-state with other likely states.
Moreover, the transition between quantum phases, arising from the addition of frustrations in a $J_1-J_2$ model, can be systematically explored using the same set of measurements.
We show that the proposed measurements have an application in producing long distance heralded entanglement between well separated quantum dots.
Relevant noise sources, such as non-zero temperatures and nuclear spin interactions, are considered.
\end{abstract}

\maketitle

\section{Introduction}

Quantum simulators~\cite{buluta2009quantum} are one of the hotly pursued topics of current quantum technology research.
Analog quantum simulators directly mimic another physical quantum system in order to explore its behaviour in greater depth.
In doing so, they provide a wide range of applications, for instance, addressing challenges in smart material design which could potentially revolutionize medicine and energy provision in the future.
While already accessible, quantum simulators will scale to much larger sizes in the near future, in doing so becoming a significant technological step on the path to full quantum computation.
A key question for such simulators is the certification of the states realized within them.
For example, simple questions such as whether the state is a genuinely quantum, pure and entangled many-body state need to be answered with available measurement schemes.
For an experimentalist who has realised a candidate state it is crucial to discriminate it from the closest classical counterpart (e.g. the Neel state for antiferromagnets), random, thermal and energetically proximal quantum states.
Here we address the question with respect to the emerging field of solid state quantum simulators~\cite{stafford_collective_1994,barthelemy2013quantum,yang_spin-state_2010,farooq_adiabatic_2015,salfi_quantum_2016,prati2012anderson}.

\begin{figure}[t]
   \centering
    \includegraphics[width=\linewidth]{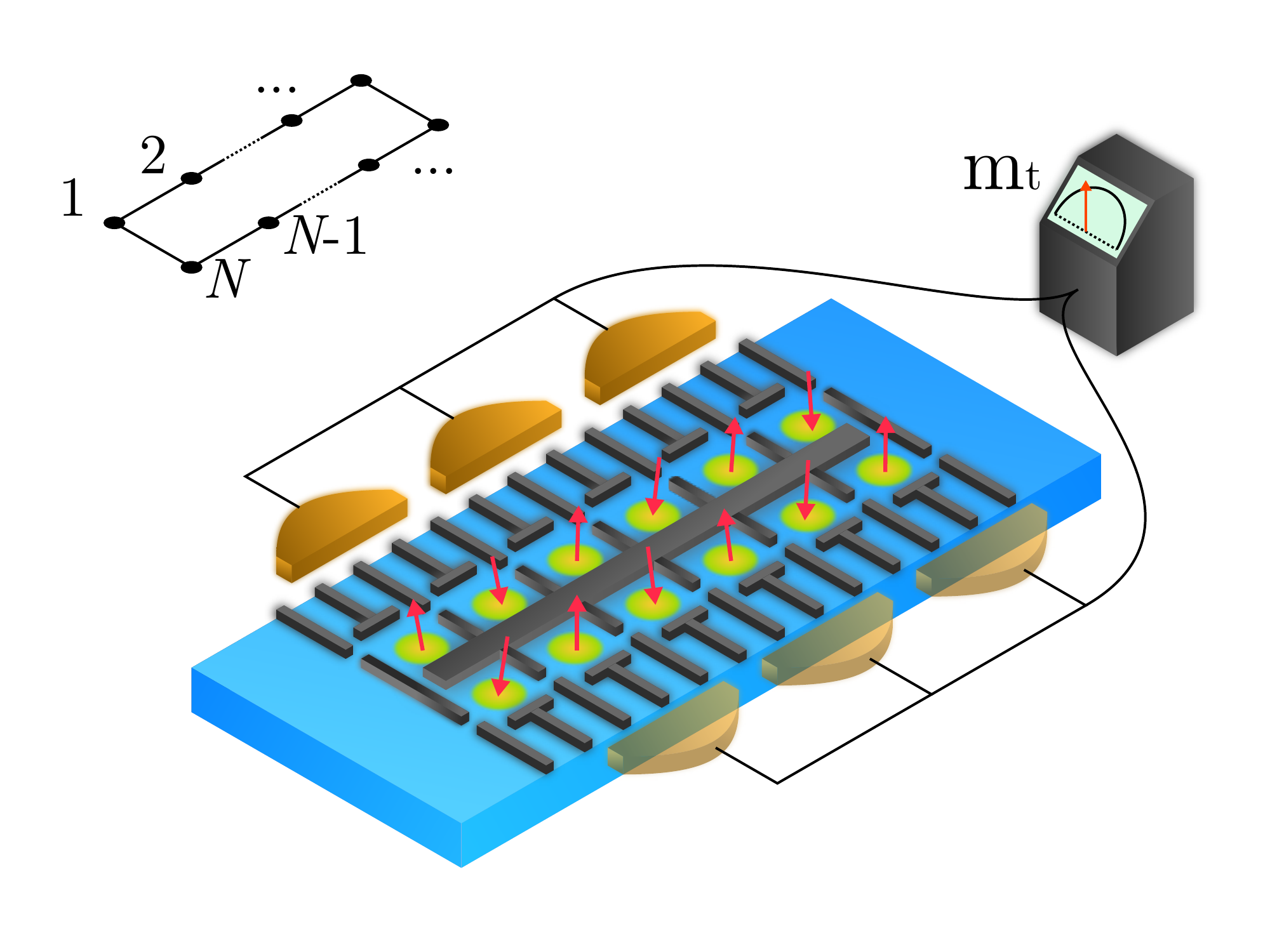}
   \caption{
   (Color online)
   \textbf{Quantum dot array spin model simulator and triplet profile readout:}
   A simplified schematic of a quantum dot array simulating a Heisenberg $N=12$ closed chain with Singlet-Triplet measurements.
   Here, grey bars represent voltage gates, each arrow a single confined electron, and the gold detectors Singlet Triplet measurements.
   These are simultaneously performed such that the total number of triplets present, $m_t$ is recorded.}
\label{fig:schematic}
\end{figure}

So far, neutral ultra-cold atoms~\cite{bloch_quantum_2012} and trapped ions~\cite{blatt_quantum_2012} have been predominantly exploited for serving as quantum simulators thanks to their high controllability and long coherence times.
Nevertheless, in order to simulate solid state systems, the presence of both particle hopping and long-range charge interactions are needed, and these are not not readily available in trapped ion and cold atom systems respectively.
Additionally, the spin exchange couplings realised in these systems tend to be small, such that any dynamics take place over long ($\sim $ ms) time-scales.
It is therefore timely, thanks to recent advances in fabrication of quantum dot arrays~\cite{puddy_multiplexed_2015}, to think about a real solid state quantum simulator.
These advances have largely been fuelled by seminal work of Loss and DiVincenzo~\cite{loss_quantum_1998}, who proposed single electron spins as qubits.
Such quantum dot arrays have also been proposed for quantum state transfer~\cite{yang_spin-state_2010} and adiabatic many-body state preparation~\cite{farooq_adiabatic_2015}.
A two-site quantum Hubbard model has been successfully simulated with dopant atoms in silicon~\cite{salfi_quantum_2016}, which are qualitatively equivalent to quantum dot arrays as far as their prospects for quantum simulations are concerned~\cite{prati2012anderson}.
Unlike cold atoms and ions, quantum dot arrays naturally have more types of interaction, such as spin-orbit~\cite{hanson2007spins}, and thus can simulate a wider range of interactions.
Moreover, their compactness allows for stronger interactions resulting in faster operations.
{
  Nevertheless, there are still challenges worth mentioning: (i) there are also strong interactions between the electrons and the environment (such as proximal nuclear spins) which decohere the simulator, and (ii) the small scale of the fabrication, and the required number of gates, makes it currently difficult to scale up to complex arrays.

Recently, Singlet-Triplet (ST) measurement in double quantum dots has emerged as the dominant tool for spin information readout.
Originally this was achieved through charge measurements~\cite{petta_coherent_2005}, motivated by decoherence free singlet-triplet qubits~\cite{levy_universal_2002}.
Radio Frequency (RF) reflectometry has since emerged as the primary method of accomplishing this~\cite{petersson2010charge,jung2012radio,basset2013single,chorley2012measuring,frey2012dipole,colless2013dispersive}.
The same measurement tool now extends beyond double dot systems to donor-dimers~\cite{house2015radio}.
These measurements discriminate between only the singlet state, and the remaining Bell-states.
Nevertheless, it is known that these measurements, in combination with particular initial states, are sufficient for universal quantum computation~\cite{rudolph_relational_2005}.
The convenience and popularity of the ST-measurements motivate us to investigate their usefulness as a tool for probing and exploiting the many-body state realized in a quantum dot array.

Independent of the physical set-up, in order to verify the performance of a quantum simulator ideally one has to fully characterise the quantum state.
The difficulty here is that by definition, a useful quantum simulator (i.e.\ with a large number of qubits) will have no exact, classically computable reference system.
Additionally, full quantum state tomography requires an exponentially large number of distinct measurements~\cite{quantum_tomography_book, gross_quantum_2010}.
Recently, there have been proposals~\cite{cramer_efficient_2010-1} for efficient tomography schemes which are applicable for those states satisfying a matrix product state ansatz, though one has to be able to perform complex multi-qubit unitary operations and measurements which are not necessarily available in the lab.

In this paper, we consider quantum dot arrays simulating the ground-state of a Heisenberg spin chain.
To characterise the state, we rely only on singlet-triplet measurements performed over nearest neighbour electron pairs, as has been experimentally demonstrated~\cite{petersson2010charge,jung2012radio,basset2013single,chorley2012measuring,frey2012dipole,petta_coherent_2005,barthel_rapid_2009,shulman_demonstration_2012}.
This allows us to build up a probability distribution over outcomes that discriminates between our target state, i.e.\ the Heisenberg ground-state, and contaminated versions.
In the presence of next-nearest neighbour interactions, realizable in recently developed multiplexed dot ladders~\cite{puddy_multiplexed_2015}, our setup can capture the quantum phase transition to a gapped, dimerized phase.
Moreover, as another application, we show that the same set of measurements can be exploited to generate heralded entanglement between distant qubits.
We investigate the performance of both applications under the influence of likely noise sources such as thermal fluctuations and hyperfine interactions with nuclear spins in the bulk.

The structure of this paper is as follows: in Section.~\ref{sec:model} we introduce the model used to describe the system and the \emph{triplet profile} that one can obtain from singlet-triplet measurements only.
In Section.~\ref{sec:characterisation} we explore the possibilities of characterizing states using these measurements only, including a quantification of how \emph{distinguishable} various states are from each other.
We demonstrate that the quantum phase transition at $J_2/J_1 \sim 0.24$ for the $J_1-J_2$ Heisenberg chain can be clearly observed.
In Section.~\ref{sec:ent_loc} we explore using singlet-triplet measurements only to \emph{localize entanglement} between two ends of an open chain.
In Section.~\ref{sec:noise} we investigate the effect of the two dominant noise sources in quantum dots --- non-zero temperature and hyperfine interactions with proximal nuclei.
Finally, in Section.~\ref{sec:experimental_realization} we propose a feasible experimental realization that could establish the validity of these methods.


\section{Model}\label{sec:model}

A key model in condensed matter physics is the Heisenberg Hamiltonian --- used in many contexts including magnetism~\cite{blundell2001magnetism}
and quantum phase transitions~\cite{sachdev2011quantum}.
It describes the interaction between $N$ spin-1/2 particles as
\begin{equation}
    H_1 = J_1 \sum_{i=1}^N \vec{\sigma}_i \cdot \vec{\sigma}_{i+1}; 
    \label{eq:ham}
\end{equation}
where $\vec{\sigma}_i=(\sigma_i^x,\sigma_i^y,\sigma_i^z)$ is a vector of Pauli operators acting on site $i$, and $J_1$ represents the nearest neighbour spin coupling.
We have assumed periodic boundary conditions, i.e. $\vec{\sigma}_{N+1}=\vec{\sigma}_1$, however, our analysis is equally applicable to open chain where increased dimerization makes the ground-state even more distinct.
We set $J_1=1$ throughout the paper, unless specified, considering it the energy scale of the system.
This anti-ferromagnetic Heisenberg model has a unique SU(2) symmetric ground-state for even lengths, $N$, known as a global singlet since it has total spin $S=0$. The lowest lying excitations are three degenerate `triplet' states, with the energy gap to these closing as $1/N$ in the limit of large $N$.

In order to simulate the ground-state of the Heisenberg Hamiltonian in a controlled way we propose a quantum dot array with exactly one electron in each quantum dot as schematically shown in FIG.~\ref{fig:schematic}(a).
A similar structure has recently been realized for multiplexing quantum dots~\cite{puddy_multiplexed_2015}.
The spin sector of the interaction between the electrons
is explained by the Hamiltonian (\ref{eq:ham}) and the coupling $J_1$ can be tuned by applying appropriate gate voltages to the gates controlling the potential barrier between neighbouring electrons.
By cooling this quantum system below its energy gap it can be initialized in its ground-state $\ket{\psi_0}$.
The central object of interest in this paper is $\ket{\psi_0}$ due to its highly entangled and non-trivial structure, described by a many-body global singlet, as well as its application for practical tasks in quantum technologies such as quantum state transfer~\cite{bose2003quantum,nikolopoulos2014quantum}.
The first stage of verifying the operation of a quantum simulator is to characterize and certify its achieved state --- hopefully the ground-state $\ket{\psi_0}$.
Ideally this could be done using full quantum tomography~\cite{nielsen2010quantum} or other more efficient methods~\cite{cramer_efficient_2010-1,gross_quantum_2010} but for quantum dot arrays a current limitation is that only Singlet-Triplet (ST) measurements on adjacent sites are feasible.
The question to be addressed here is to what extent characterisation and certification of a state is possible under this restriction.

The ST-measurement can be described by the following projectors 
\begin{eqnarray}\label{Projections}
    \mathbf{P}_s &=& \ket{\psi_-}\bra{\psi_-}, \cr
    \mathbf{P}_t &=& \boldsymbol{1} - \mathbf{P}_s,
\end{eqnarray}
where $\ket{\psi_-} =\frac{1}{\sqrt{2}}\left(\ket{\uparrow\downarrow} - \ket{\downarrow\uparrow}\right)$ is the singlet with $\ket{\uparrow}$ and $\ket{\downarrow}$ representing spin up  and down respectively.
If the quantum simulator operates perfectly, i.e.\ initializes in the ground-state $\ket{\psi_0}$, then thanks to the SU(2) symmetry of the system the reduced density operator of any pair spin qubits will be a Werner state~\cite{oconnor_entangled_2001}
\begin{equation}
    \rho = \alpha \mathbf{P}_s + (1 - \alpha)\dfrac{\mathbf{P}_t}{3},
\end{equation}
with $0 \leq \alpha \leq 1$.
In this sense ST-measurements are picked out as a preferred `basis' for all SU(2) symmetric states.

Let's assume that the system is described by the density matrix $\rho$, ideally $\ket{\psi_0}\bra{\psi_0}$.
Performing ST-measurements on all $N/2$ consecutive pairs of spins, i.e. qubits $(1,2)$, $(3,4)$, \ldots, $(N-1,N)$, results in $2^{N/2}$ different outcomes according to the singlet or the triplet output of each measurement.
For example in a chain of length $N=4$ any of the outcomes $ss$, $st$, $ts$ or $tt$ may occur with a certain probability.
For any string of outcomes $x=x_1x_2...x_{N/2}$ (with each $x_i$ being $s$ or $t$) the total projection operator is
\begin{equation}
\Pi_x = \bigotimes_{i=1}^{N/2} \mathbf{P}^{2i-1,2i}_{x_i}
\end{equation}
where $\mathbf{P}^{2i-1,2i}_{x_i}$ are the same projectors as in Eq.~(\ref{Projections}) acting on qubits $2i-1$ and $2i$.
Thus, the probability of getting the string $x$ as the outcome of the measurements is $\Tr(\Pi_x \rho)$.
For example the probability of getting the result $x=stts$ for a $N=8$ state is $\Tr(\mathbf{P}_s^{12} \mathbf{P}_t^{34} \mathbf{P}_t^{56} \mathbf{P}_s^{78} \rho)$.
We can further compress the number of outcome results by grouping together all result strings featuring the same number of measured triplets, thus creating a \emph{triplet profile}:
\begin{equation}
    p(m_t) = \sum_{x \in X_m} \Tr(\Pi_x \rho)
    \label{eq:triplet_profile}
\end{equation}
where $X_m$ denotes the set of all result strings with \emph{exactly} $m$ triplet occurrences.
This yields a concise characterisation of a state that is both easy to measure experimentally and to compute numerically.
Performing the sum in Eq.~(\ref{eq:triplet_profile}) loses all information about how `grouped' triplet excitations are, nevertheless, a surprising amount information can be gleaned from $p(m_t)$, including features heralding many-body entanglement.
For example, one such feature that arises is that $p(m_t=1)=0$ for all global singlets.
This arises from their spin-0 nature --- they can have no overlap with the spin-1 subspace which includes all configurations of a single triplet.
Indeed, under the reasonable restriction of translational invariance, classical states can only ever produce a binomial distribution for $p(m_t)$, and any deviations such as oscillations herald entanglement.

\section{Characterization of Simulator\label{sec:characterisation}}

\begin{figure}[ht]
    \centering
     \includegraphics[width=\linewidth]{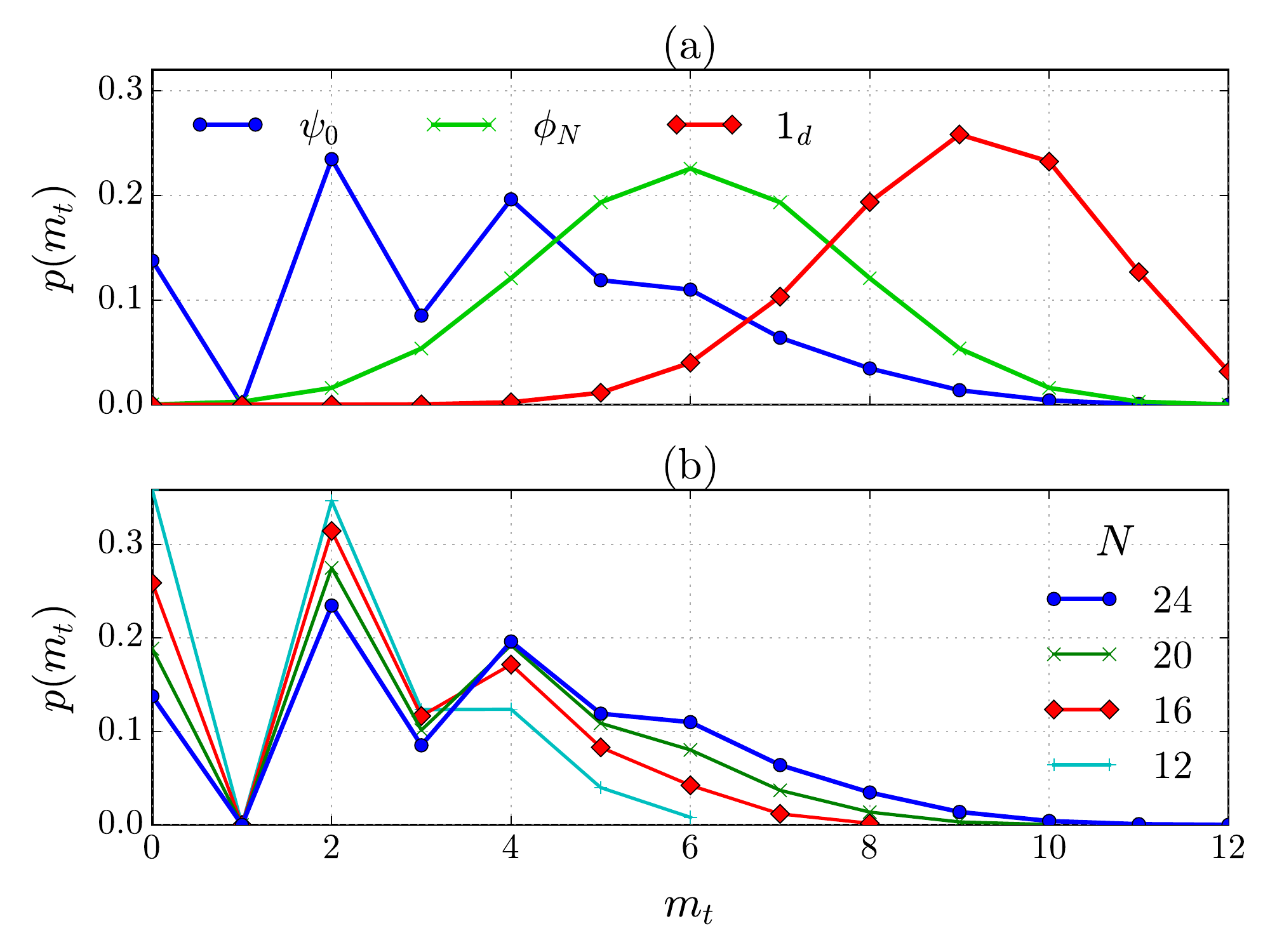}
    \caption{
    (Color online) \textbf{Discrimination of the ground, classical, and random states: (a)}
    Triplet probability profiles for a number of states, namely the anti-ferromagnetic Heisenberg ring ground-state, $\psi_0$, the classical anti-ferromagnetic Neel state, $\phi_N$, and the normalised identity, $\mathbf{1}_d$, all of size $N=24$ (thick lines).
    \textbf{(b)}
    Scaling of the triplet profile with $N$ for $\psi_0$.}
\label{fig:certification_basic}
\end{figure}

In order to characterize the quantum state of the simulator we first calculate the full triplet profile $p(m_t)$ of the ground-state $\ket{\psi_0}$
and other likely states which may occur due to imperfections or malfunctioning of the quantum simulator.
In particular, we consider the classical Neel state $\phi_N = \ket{\uparrow\downarrow\uparrow\downarrow\uparrow\downarrow\dotsc}$ and the maximally mixed state $\mathbf{1}_d$, which represents an infinite temperature thermal state.
In FIG.~\ref{fig:certification_basic}(a) we plot the triplet profile $p(m_t)$ as a function of the number of triplet occurrences $m_t$ for a chain of length $N=24$ for all the three states.
As can be easily calculated, $\phi_N$ and $\mathbf{1}_d$ are both characterised by binomial distributions centred on $\frac{1}{2}$ and $\frac{3}{4}$ respectively, whereas $\ket{\psi_0}$ produces a highly non-trivial oscillatory shape.
For example, the zero-probability $p(m_t=1)$ dip is very prominent, and also forms part of an oscillatory structure between odd and even occurrences of triplets.

The scaling of $p(m_t)$ for $\psi_0$ with size of system $N$ is also shown in FIG.~\ref{fig:certification_basic}(b) --- one can see that overall the features change slowly, with the average $m_t$ increasing with $N$ under the `oscillating' envelope.
As such, although the first `fringe' contrast decreases with $N$ slightly, the second increases and so on such that they should not be washed out in the thermodynamic limit.

\begin{figure}[ht]
    \centering
     \includegraphics[width=\linewidth]{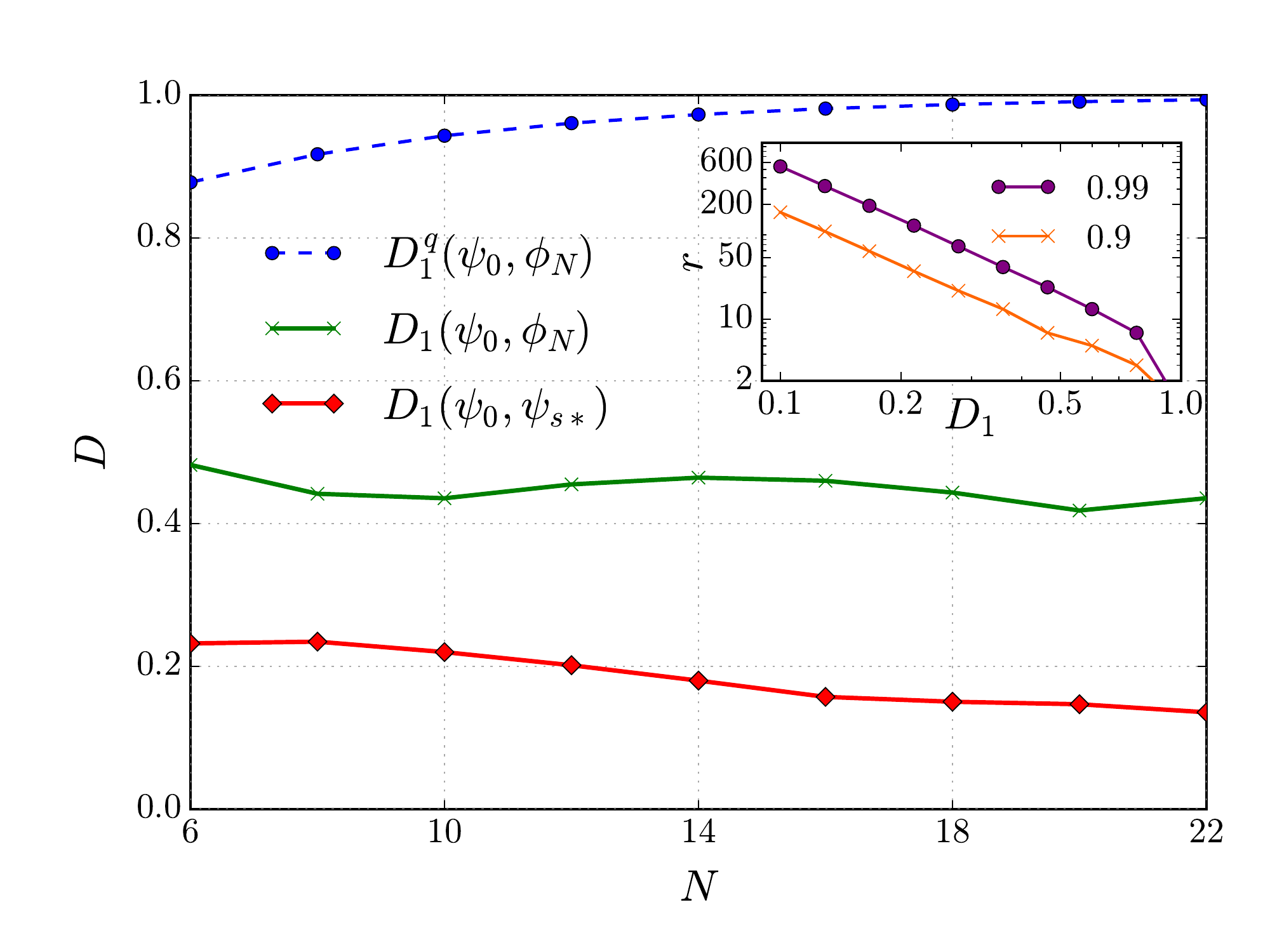}
    \caption{
    (Color online)
    \textbf{Comparison of the single shot distinguishability under an optimal measurement vs.\ triplet profile measurements:}
    Scaling with system size of the single shot distinguishability between Heisenberg ground-state $\psi_0$, first excited singlet state $\psi_{s*}$, and the Neel state $\phi_N$. Full lines denote distinguishability under triplet profile measurements, whereas dashed lines denote the full quantum distinguishability $D^q_1$. Note that $D^q_1(\psi_0, \psi_{s*})$ is not shown since it is always 1 --- the states being orthogonal.
    \textbf{Inset:} number of repeat measurements, $r$, required to distinguish two states with probability ($0.9$, $0.99$) for varying $D_1$.}
\label{fig:distinguishability_scaling}
\end{figure}

Full quantum tomography is usually very demanding either in terms of sheer number of measurements or the complex many-body basis of such operations.
Instead, we wish quantify the extent to which our ST-measurements can distinguish between likely quantum states (i.e. selected based on some prior intuition).
A fundamental quantity here is what we shall call the single-shot-distinguishability, $D_1$, which quantifies the advantage a single measurement gives when guessing between two equally probably states such that the overall chance of success is $\frac{1}{2}\left(1 + D_1 \right)$.
If $D_1=0$ then the measurement yields no information at all about which state is present, whereas if $D_1=1$ it perfectly discriminates them.
If a measurement gives rise to two possible probability distributions, $p_1(a)$ and $p_2(a)$, over outcomes $a$ then $D_1$ is given by~\cite{baigneres2004far}:
\begin{equation}
	D_1 = \frac{1}{2}\sum_a \left| p_1(a) - p_2(a) \right|
	\label{eq:single_shot_distinguishability}
\end{equation}
which essentially formalizes the strategy of guessing whichever state is more likely to give result $a_i$ each time.
It has been shown~\cite{helstrom_quantum_1969} that for two quantum states that $\rho$ and $\sigma$ the maximum distinguishability is given by:
\begin{equation}
	D^q_1 = \frac{1}{2}\left\| \rho - \sigma \right\|_{\text{tr}},
\end{equation}
where $\left\|A\right\|_{\text{tr}}=Tr(\sqrt{AA^\dagger})$ is the trace norm.
It is worth mentioning that the optimal measurement needed to yield $D_1^q$ is likely to be a globally entangled projective measurement that is again not feasible.

An important aspect to investigate is whether the triplet profile's ability to distinguish scales well with system size.
In FIG.~\ref{fig:distinguishability_scaling} we present the single shot distinguishability between two states under both a triplet profile measurement, $D_1$, and an optimal quantum measurement, $D_1^q$, as it scales with system size $N$.
First consider the case of $\psi_0$ and $\phi_N$ --- $D^q_1$ rises to $1$ with $N$ while $D_1$ for the triplet measurement hovers at just under half this, with possibly a slight decrease with $N$.
In this sense, a value of $D_1 \sim 0.45$ is decent.
As an illustration of two states that are almost worst-case scenario, we also present the distinguishability of the ground-state $\psi_0$ and the first excited global singlet, $\psi_{s*}$, which represents the smallest energy, symmetry preserving excitation that could occur.
Clearly these two states are orthogonal and thus $D^q_1(\psi_0, \psi_{s*})=1$, but in character they are very similar.
Nonetheless the triplet profile produces a non-zero distinguishability, as can be seen from FIG.~\ref{fig:distinguishability_scaling}, which also only decreases slowly with $N$ --- not surprising since these two states are becoming closer relatively within the Hilbert space.
To give a sense of what these values of $D_1$ mean in practice, the inset of FIG.~\ref{fig:distinguishability_scaling} shows the number of required measurements, $r$, in order to achieve a total probability of successfully distinguishing two states, given the naive strategy of guessing independently which state was present each repeat .
This sub-optimal scheme casts the overall distinguishability as that of between two binomial distributions.
For example, if we take $D^q_1(\psi_0, \phi_N) \sim 0.43$, then $27$ measurements would be required to guess which state was present with 99\% success,
  as shown in the inset of Fig.~3.

\section{Quantum Phase Transition in the $J_1-J_2$ Model \label{sec:characterisationj1j2}}

\begin{figure}[ht]
    \centering
     \includegraphics[width=\linewidth]{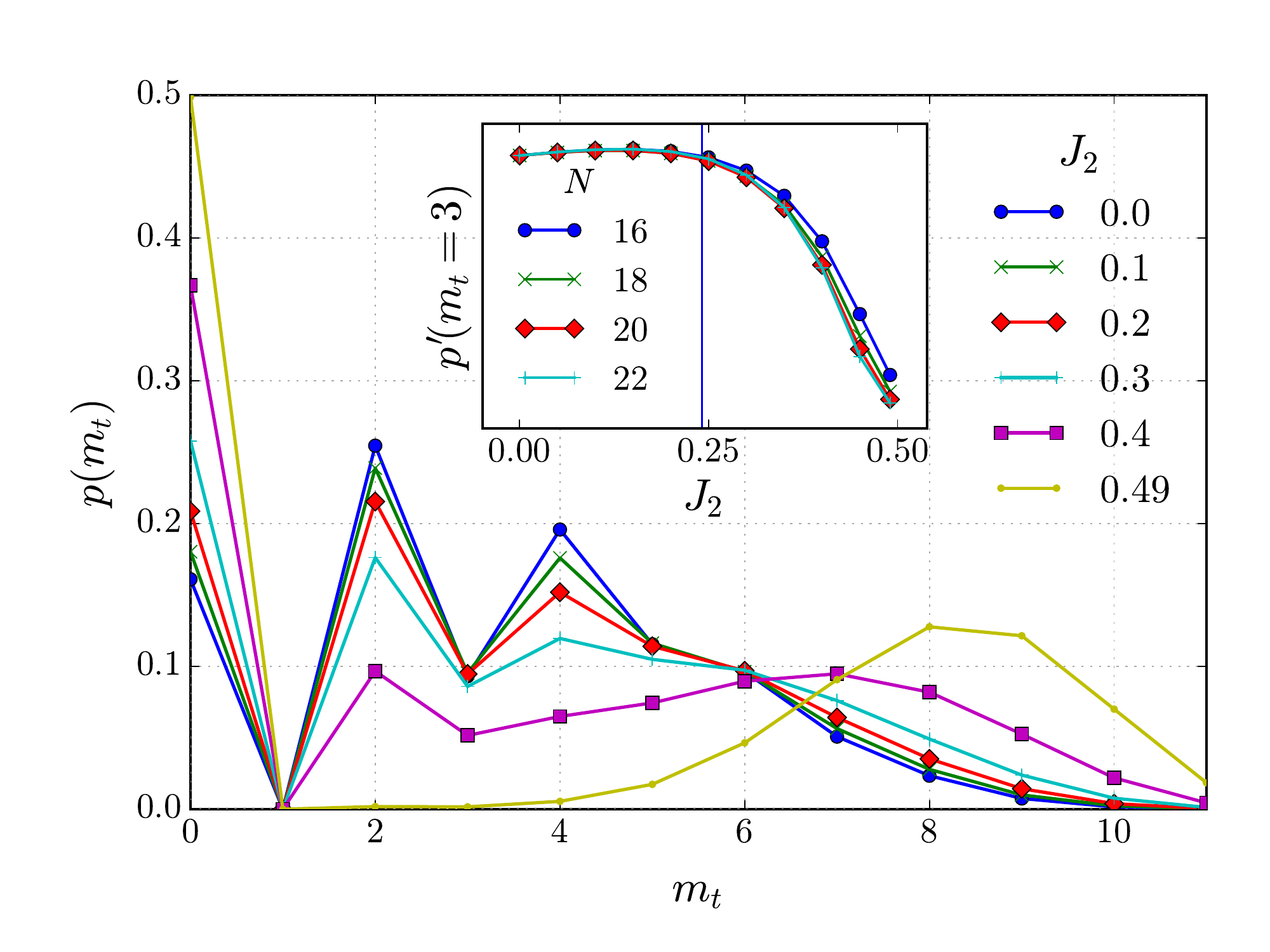}
    \caption{(Color online)
    \textbf{Observing the $J_1-J_2$ quantum phase transition with the triplet profile:}
    Triplet profile for the $N=22$ Heisenberg ring ground-state, $\psi_{J_2}$, across the $J_1-J_2$ phase transition.
    \textbf{Inset:} normalised probability of measuring three triplets varying with $J_2$ for various $N$.
    The vertical line denotes the exact critical point.}
\label{fig:certification_phase_transition}
\end{figure}

In some condensed matter systems long range interactions are not negligible and play a crucial role in the character of the system.
The simplest example is  the $J_1-J_2$ model with the Hamiltonian
\begin{equation}
H_2= J_1 \sum_{i=1}^N \vec{\sigma}_i \cdot \vec{\sigma}_{i+1} + J_2 \sum_{i=1}^{N-1} \vec{\sigma}_i \cdot \vec{\sigma}_{i+2},
\label{eq:j1j2model}
\end{equation}
where $J_2$ is the next nearest neighbour coupling strength.
This model exhibits a quantum phase transition from a gap-less Heisenberg phase to a gapped dimerized phase through increasing $J_2$.
This infinite order quantum phase transition, in the Berezinskii–--Kosterlitz–--Thouless universality class, happens at $J_2/J_1 \simeq 0.24$.
No standard quantities behave non-analytically across the transition, and instead properties such as the ground-state `fidelity susceptibility' ~\cite{thesberg2011general} or excited state fidelity~\cite{chen_fidelity_2007} must be used to locate the critical point.
Another interesting point in the dimerized phase is the Majumdar-Ghosh point at $J_2/J_1=0.5$.
Here the ground-state is fully dimerized and can be explained as an equal superposition of $\bigotimes_{i=1}^{N/2} \ket{\psi_-}$ and its equivalent, but one site translated, form.
The model in Eq.~\ref{eq:j1j2model} could well be realised in future quantum dot arrays via a `zig-zag' ladder geometry.

In FIG.~\ref{fig:certification_phase_transition} we show the triplet profile for the ground-state of $H_2$ for a number of $J_2$ values across the phase transition.
As $J_2$ approaches the Majumdar-Ghosh point (i.e. $J_2/J_1=0.5$) the structure of an equal superposition of dimerizations becomes apparent --- half of the state is exactly singlet pairs aligned with the measurements, and the other half appears as the identity since it is singlet pairs between the measurements.
We note also that the rate and quality of change is different on either side of the critical point.
This is shown more clearly in the inset of FIG.~\ref{fig:certification_phase_transition}, where only the probability of getting three triplets (i.e. $m_t=3$) is plotted versus $J_2$.
In order to have a better perception of the effect of length $N$ we have normalized the probabilities to the $J_2=0$ case, $p'(m_t=3)$, for various lengths.
Finally, this effect is not limited to $p(m_t=3)$ --- other values and combinations of $m_t$ also give the same behaviour.
However, due to the continuous nature of the transition, no such quantities are expected to show very sharp features, especially for short chain lengths.

\section{Heralded entanglement of distant spins\label{sec:ent_loc}}

We now show a potential quantum information application of using solely singlet triplet measurements in the form of \emph{long-distance} entanglement.
Generating perfect entanglement over arbitrary distances will likely be required for many quantum information tasks.
In a many-body system, it has been shown~\cite{verstraete_entanglement_2004,popp_localizable_2005}, that performing measurements on part of a system can localize entanglement between the remaining, unmeasured parts.
This is known as \emph{localizable entanglement}, and we demonstrate here that singlet-triplet measurements on the ground-state of the Heisenberg chain can probabilistically localize entanglement between any two qubits.
For applicability, to a quantum bus for example, we consider now an \emph{open} chain, where the first and last quantum dots are desired to be entangled.
The nature of global singlet states guarantee that if all but one pair of spins is measured and found in the singlet state, the final pair must also be in the singlet state.
As previously described, this follows from the fact that all $m_t=1$ states have spin-1, and no overlap with the SU(2) subspace.
The generation of a perfectly entangled singlet is therefore reliant on the probability of finding this all-singlet outcome, $q(m_t=0)$, but is certain to be there (i.e. \emph{heralded}) if the measurement succeeds.
Note that we use the symbol $q(m_t)$ for the probability of finding $m_t$ triplet outcomes in our ST-measurements for the heralded entanglement scheme, which leaves one pair unmeasured, to discriminate it from $p(m_t)$ in the previous section in which all qubits are measured.
{
  Compared to a dynamic, gate based-scheme, the simultaneous nature of the measurement minimizes the time required and thus exposure to decoherence.

Since any ground-state with SU(2) symmetry displays this feature, we can also think about engineering the exchange coupling strengths along the chain to promote this configuration.
One option is to weaken the coupling of just the end spins as
\begin{equation}
	H_e=J_e(\vec{\sigma}_1\cdot \vec{\sigma}_2 + \vec{\sigma}_{N-1} \cdot \vec{\sigma}_{N}) + J_1 \sum_{i=2}^{N-2} \vec{\sigma}_{i} \cdot \vec{\sigma}_{i+1}
\end{equation}
where $J_e$ is the ending coupling and is smaller than $J_1$.
For $J_e \ll J_1$ it is known that a very high entanglement is established between the outermost spins in the ground-state of the system~\cite{campos_venuti_qubit_2007}.
However, this entanglement is thermally unstable due to a vanishing energy gap.
We combine this scheme, using larger values of $J_e$, and a heralding ST-measurement to achieve perfect entanglement with a higher rate.
As another way to improve the probability $q(m_t=0)$ we may also consider a Hamiltonian with alternating couplings as
\begin{equation}
H_a= J_1 \sum_{i=1}^{N-1} [1-(-1)^{i}\delta] \text{  } \vec{\sigma}_{i} \cdot \vec{\sigma}_{i+1}
\end{equation}
where $\delta$ is the dimensionless anisotropy parameter.

In FIG.~\ref{fig:ent_loc_basic}, we show the probability of an all singlet result, $q(m_t=0)$, for these three cases, as the length of chain, $N$, varies for the case of $J_e=0.5J_1$ and $\delta=0.1$.
Weakening the end-bonds yields a consistent improvement in long-distance entanglement over the normal Heisenberg chain, but both still decrease exponentially with length.
The heralded nature of the entanglement means that for small enough chains repetition could still make the procedure viable.
Moving to the ground-state of the $H_a$, we find that $q(m_t=0)$ becomes almost constant with $N$ at a value of $\sim0.3$.
  A subtlety here is that engineering a Hamiltonian in this way can reduce the size of the energy gap, making the ground-state harder to prepare.
  This is in fact the case for both $H_e$ and $H_a$ above.
  Although this means that reaching the ground-state via direct cooling becomes more difficult, adiabatic state preparation has been shown to much alleviate the issue~\cite{farooq_adiabatic_2015}.


\begin{figure}[ht]
    \centering
     \includegraphics[width=\linewidth]{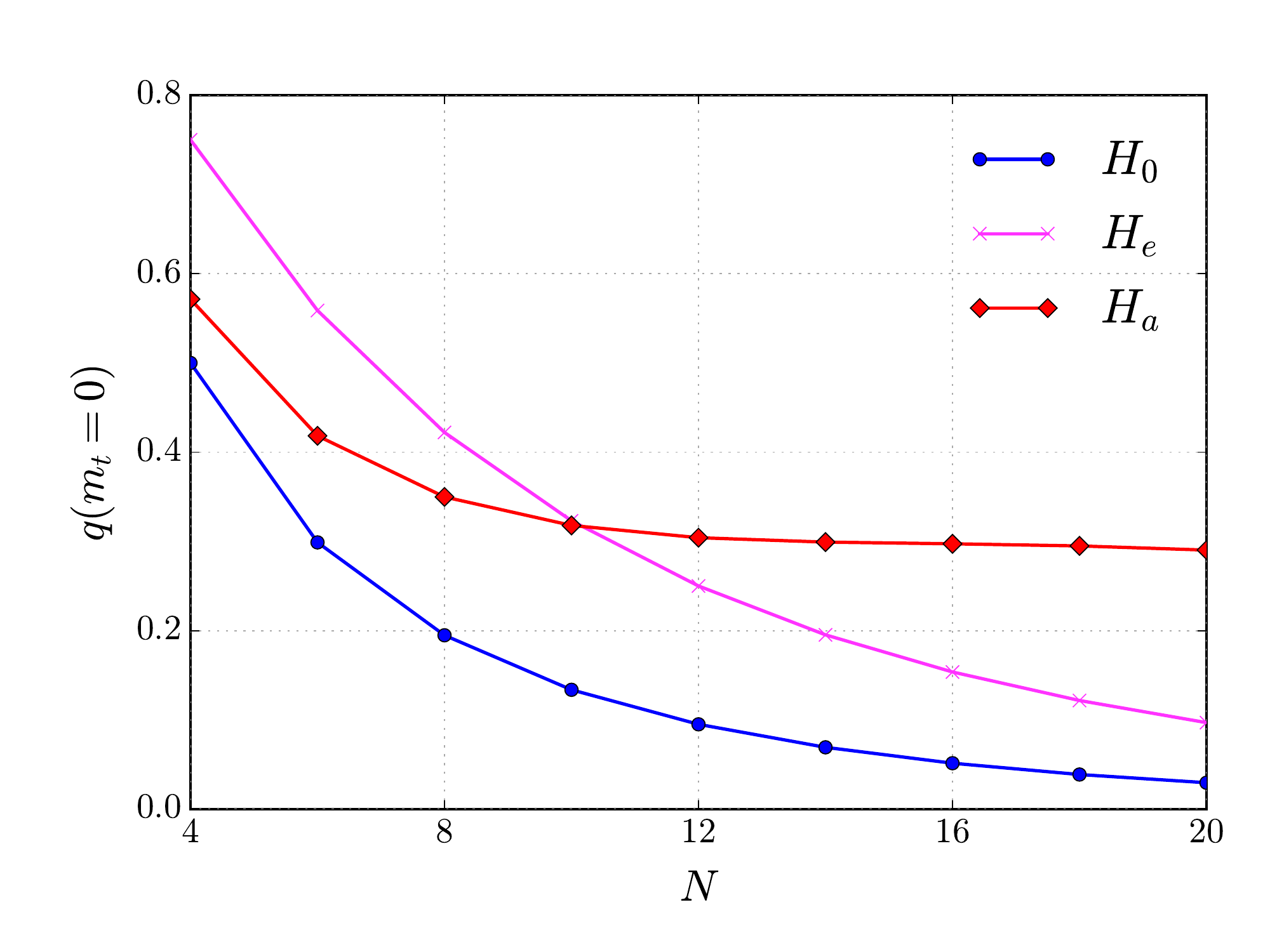}
    \caption{
    (Color online)
    \textbf{Localizing entanglement to the ends of a chain using singlet-triplet measurements on the middle $N-2$ qubits:}
    Probability of finding all singlets after measuring the middle $N-2$ spins of the ground-state of an open Heisenberg chain with various coupling configurations.
    From this outcome, a perfect singlet in the remaining two spins at either end is heralded.
    The three Hamiltonian configurations are: $H_0$  --- constant coupling, $H_e$ --- end couplings weaker by 50\%, $H_a$ --- alternate couplings weaker by 20\%.}
    \label{fig:ent_loc_basic}
\end{figure}

Finally, we point out that if full Bell-state measurements are possible on nearest neighbour spins, then perfect entanglement is \emph{always} achieved between the ends.
The state is not always the singlet Bell-state, but can be identified or corrected simply by counting the number of each Bell-states found and requiring the whole state to still be spin-0.
This is essentially the same mechanism as addressed in~\cite{barjaktarevic_measurement-based_2005}.

\section{Imperfections} \label{sec:noise}

The goal of our simulator is to create the ground-state of the Heisenberg Hamiltonian.
In reality, thermal fluctuations spoil the quantum state of the system resulting in a thermal state
\begin{equation}
\rho^\text{th}(\beta) = \frac{e^{-\beta H}}{\Tr(e^{-\beta H)}}
\label{eq:thermal_state}
\end{equation}
where $\beta=1/k_B T$ and $k_B$ denotes the Boltzmann constant.
Performing the characterization ST-measurements on a thermal state result in a triplet profile $p(m_t)$ which is shown in FIG.~\ref{fig:noise_fld}(a). From the figure, we find that up to $k_B T / J_1=0.2$ (approximately the gap of the Hamiltonian) the observed triplet profile is largely unchanged.
Between $k_B T / J_1=0.5$ and $k_B T / J_1=1$ the oscillations suggesting many-body entanglement die out, and above the state appears largely classical.
Since we know that for temperatures smaller than the energy gap the thermal state has very close to unit fidelity with the ground-state, what the result in Fig.~\ref{fig:noise_fld}(a) shows is that our singlet-triplet profile is sensitive to any rise in temperature that would significantly affect the state.
Although we can positively identify a departure from the ground-state in this way, attributing the noise specifically to thermal fluctuations or identifying the temperature poses a greater challenge, though one worth investigating in the future.

Another dominant form of noise\cite{taylor_relaxation_2007} arises from each electron's hyperfine interaction with proximal nuclei.
This manifests as an isotropic, normally distributed random static magnetic field for each site, which we can model with the Hamiltonian
\begin{equation}
H^\text{nuc}(B_n) = J_1 \sum_{i=1}^{N-1} \vec{\sigma}_{i} \cdot \vec{\sigma}_{i+1}  +  \sum_i^N \mathbf{B}_i \cdot \mathbf{\sigma}_i
\label{eq:bnuc}
\end{equation}
where $\mathbf{B}_i$'s are effective magnetic fields with random directions.
The amplitude of these fields are determined by a Gaussian probability distribution as
\begin{equation}
P(\mathbf{B}) = \frac{1}{(2\pi B_{n})^{3/2}} \exp\left(  -\frac{\mathbf{B}\cdot\mathbf{B}}{2B_{n}^2} \right)
\end{equation}
where $B_n$ is the variance of the distribution and quantifies the strength of the nuclear field noise.
The noise is quasi-static (changes slowly relative to the electron dynamics) and thus we can think of each experimental run as having a fixed set of random fields and simply average over many runs until convergence is reached.

\begin{figure}[ht]
    \centering
    \includegraphics[width=\linewidth]{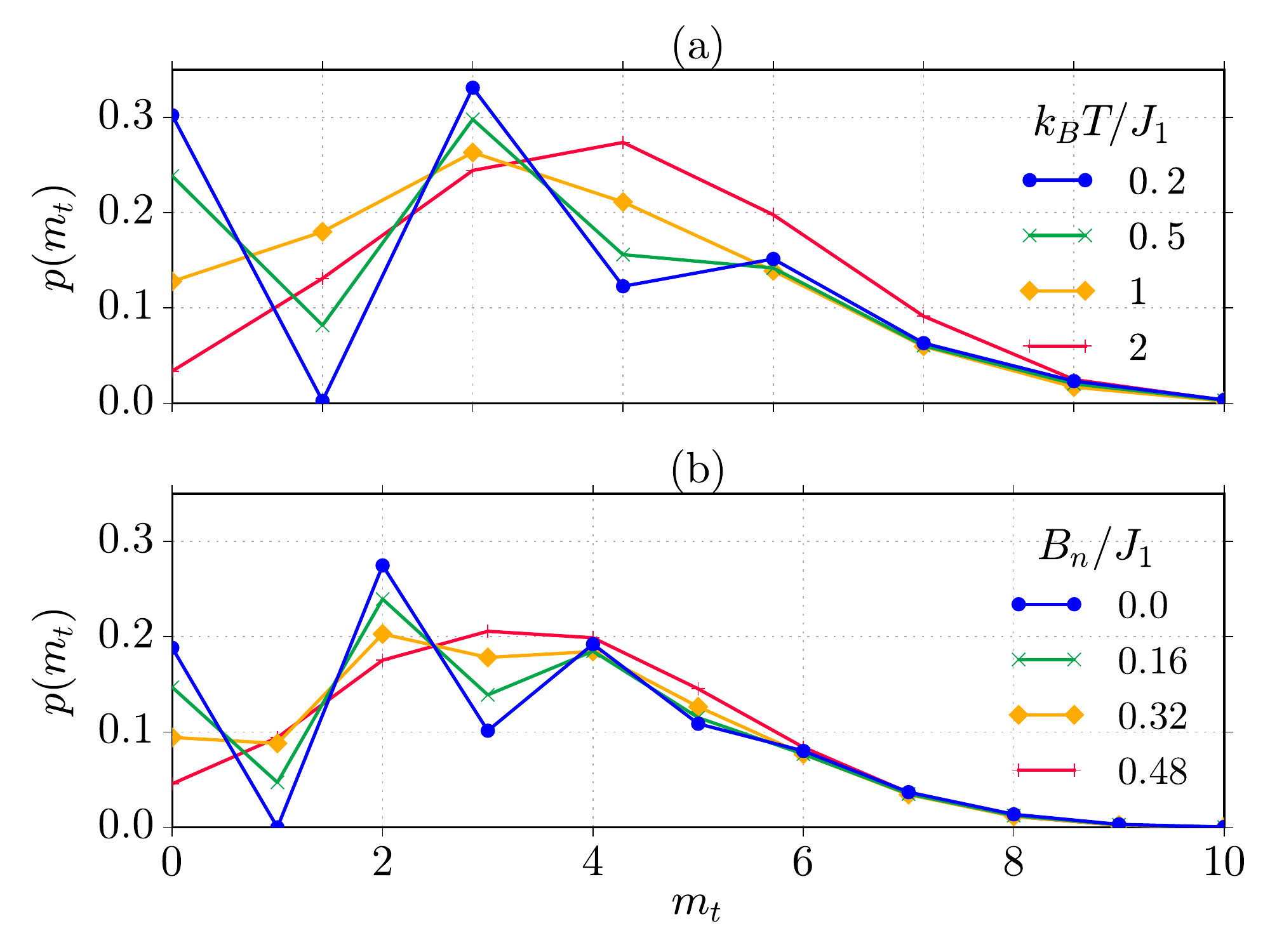}
    \caption{(Color online)
      \textbf{
        Characterising the Heisenberg chain under the influence of temperature and hyperfine interactions: (a)}
      Triplet profile of the Heisenberg ring ($H_0$) ground-state for varying temperature $T$, here for chain length $N=14$.
    \textbf{(b)}
    The same but for varying random nuclear field strength $B_n$, here for chain length $N=20$.
    It is worth mentioning that a realistic (but pessimistic) estimation of the hyperfine interaction is $B_\text{n} / J_1 \sim 0.1$~\cite{petta_coherent_2005}.
    This will have little effect on state discrimination.
    }
    \label{fig:noise_fld}
\end{figure}

In FIG.~\ref{fig:noise_fld}(b), we find that the nuclear noise quickly changes the triplet profile such that $B_n<0.1$ would likely be required for a decent characterisation.
Above $B_n \sim 0.3$ the oscillations disappear.
Actual values for the bare value of $B_n / J_1$ estimate it below 0.1~\cite{petta_coherent_2005}, which hardly affects our triplet profile characterisation.
  Moreover, many successful avenues exist for reducing the effect of the nuclear noise, such as dynamical decoupling~\cite{malinowski2016notch}, and moving to Si/SiGe quantum dots~\cite{kawakami2014electrical}, though these both introduce their own challenges for scaling to dot array simulators.

\begin{figure}[]
    \centering
     \includegraphics[width=\linewidth]{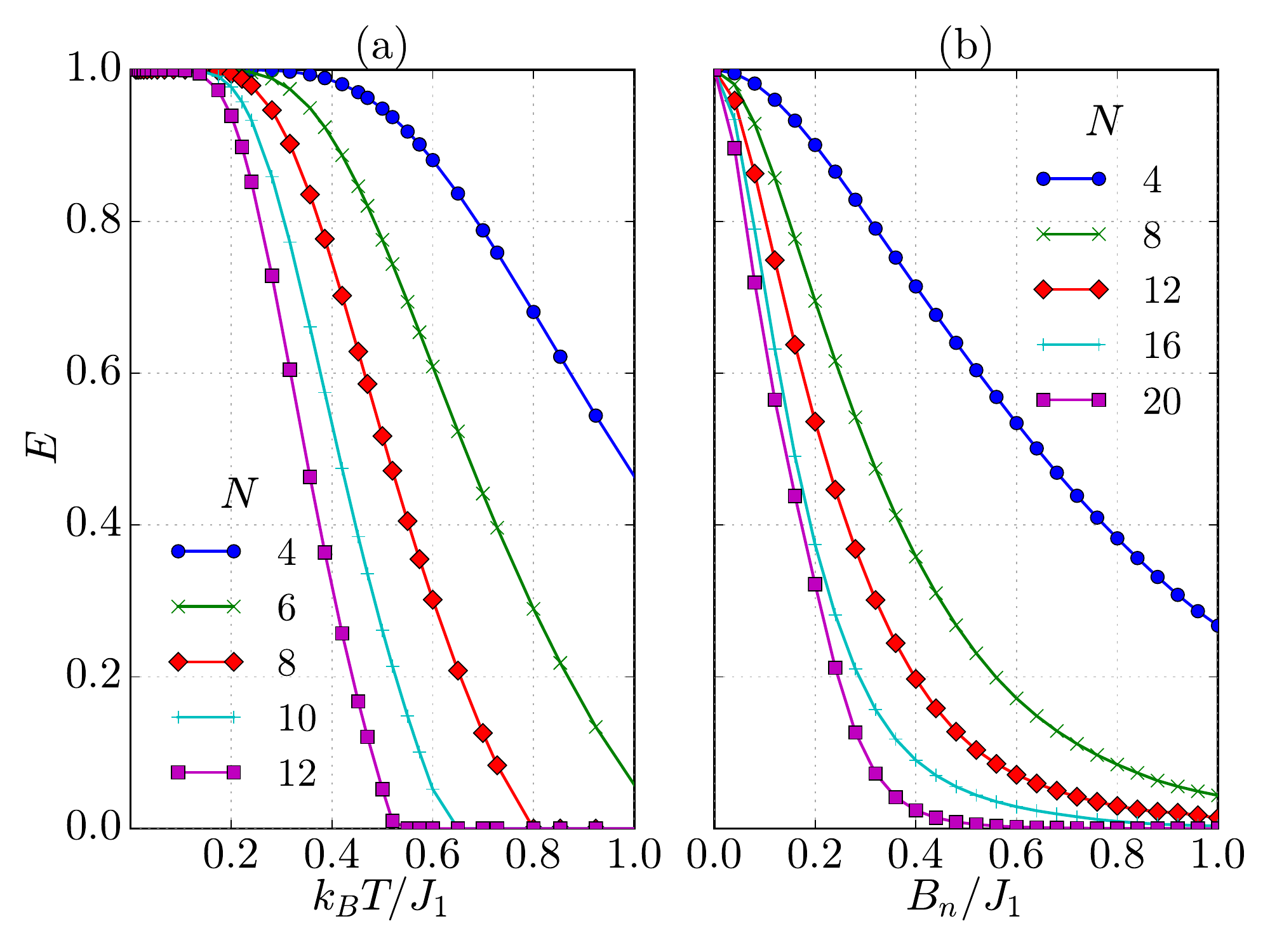}
    \caption{
    (Color online)
    \textbf{Entanglement localization to the ends of an open chain under the influence of temperature and hyperfine interactions: (a)}
    Entanglement (as measured by concurrence) between the two furthest spins after an all-singlet measurement result on the remaining middle section of the chain, varying with temperature $1/\beta$ and length $N$.
    \textbf{(b)}
    The same, but now varying with the strength of the random nuclear field $B_n$. It is worth mentioning that a realistic (but pessimistic) estimation of the hyperfine interaction is $B_\text{n} / J_1 \sim 0.1$~\cite{petta_coherent_2005}. For all considered lengths, this yields high entanglement.}
    \label{fig:noise_ent_loc}
\end{figure}

The effect of noise on long distance entanglement could be two-fold, it could change the probability of getting an all-singlet measurement, $q(m_t=0)$, and it could also make the resultant state shared between the end qubits less entangled.
In practice we find that $q(m_t=0)$ is roughly constant across the region of interest for both temperature variation and hyperfine interaction.
So, it suffices to consider only the \emph{remaining} entanglement, $E$, which we characterise with the concurrence~\cite{hill_entanglement_1997} on the reduced density matrix of the ends post-measurement.

In FIG.~\ref{fig:noise_ent_loc}(a) we show how this remaining entanglement varies as a function of temperature.
As the figure shows, there is a plateau of low temperature for which perfect entanglement still remains, though this drops with $N$ and can be again be linked with the Hamiltonian's gap.
Similarly to the case of state characterization, we find that nuclear noise has a much more immediate effect on the long-distance entanglement
rather than on $q(m_t)$.
In FIG.~\ref{fig:noise_ent_loc}(b) we plot entanglement versus $B_n$.
As the figure shows for $B_n<0.1J_1$, which as mentioned is a conservative estimation based on experiment~\cite{petta_coherent_2005}, the entanglement remains high even for chains as long as $N=20$.

  Another potential source of error in quantum dot array simulators are fluctuations in the charge potential landscape.
  The overall effect can be modelled to first order as a random fluctuation of $J_1$ about its mean value~\cite{wardrop2016characterization}.
  Since this type of noise maintains the SU(2) symmetry of the system, the essential arguments regarding oscillations in the triplet profile and localizing heralded entanglement remain intact.
  Indeed, there is evidence that the overall groundstate of a system with moderately random couplings is very similar in terms of character and utility~\cite{farooq_adiabatic_2015,petrosyan2006coherent,petrosyan2010state}.
  In Fig.~\ref{fig:fld_rand_js} we show the effect of this noise on state characterization as well as the average fidelity between the ideal ground-state and many realizations of the erroneous ground-state, $\bar{f}(\psi_0, \psi_0^{\sigma_{J}})$.
  One can see that the average fidelity remains above 85\% for $\sigma_J < 0.1 J_1$, which is a high level of randomness.
  The corresponding change in the triplet profile also becomes noticeable with increasing $\sigma_J$, and as expected, $p(m_t=1)$ remains zero throughout.
  Since even with this noise, the conditions for entanglement localization using singlet-triplet measurements are met, that scheme in its basic form is not affected.
  One observation is that the slight randomization of $J_1$ actually on average raises the chance of perfect entanglement, $q(m_t=0)$, when compared to a Heisenberg chain (data not shown).

  A final source of potential error worth discussing is the singlet-triplet readout fidelity, which for a RF-reflectometry based method reduces to the error in distinguishing two levels of capacitance.
  We have assumed this readout to be perfect throughout, for two main reasons.
  Firstly, this measurement is already very sensitive~\cite{ares2016sensitive} in comparison to the other sources of error.
  Secondly, our measurement is single-shot, which means that its sensitivity can be increased simply by extending the integration time.

\begin{figure}[]
  \centering
  \includegraphics[width=\linewidth]{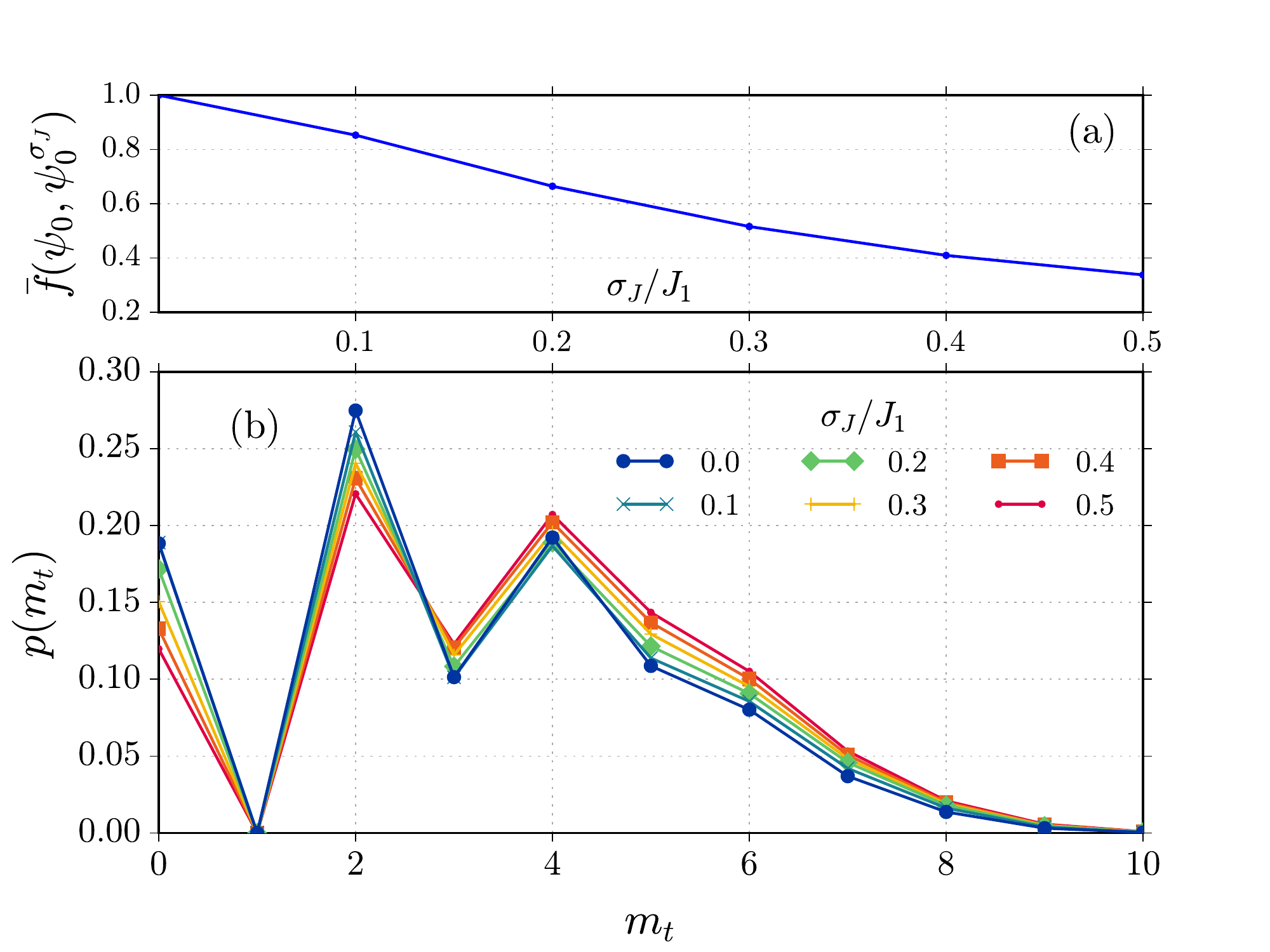}
  \caption{
    (Color online)
    \textbf{Effect of random couplings caused by charge fluctuations. (a)}
    Average fidelity of the ideal groundstate, $\psi_0$, with many realizations of the erroneous ground-state generated due to random coupling noise, $\psi_0^{\sigma_J}$, as a function of the strength of those fluctuations, $\sigma_J$, for chain length, $N=20$.
    \textbf{(b)}
    Averaged triplet profile for the ground-state of a Heisenberg chain with fluctuating couplings, varying with the strength of that fluctuation, $\sigma_J$.
  }
  \label{fig:fld_rand_js}
\end{figure}

\section{Experimental Realization\label{sec:experimental_realization}}

In this section, we discuss a potential experimental realisation.
A SEM image of a gate-defined dot array, recently developed in~\cite{puddy_multiplexed_2015}, is shown in Fig.~\ref{fig:schematic2}(a) in which fourteen quantum dots interact in a $2\times 7$ array.
Similar structures are being developed in other groups~\cite{baart2016nanosecond,petersson2009microwave}.
The ladder structure, shown in Fig.~\ref{fig:schematic2}(a), is capable of realizing a $N=14$ open chain, but in principle a ring geometry is possible and both yield qualitatively similar results.
Although the exchange coupling $J_1$ can be tuned to very large values, due to the limitations imposed by electronics speeds, a $\sim 1 \text{GHz}$ value is preferable.
In fact, in Ref.~\cite{petta_coherent_2005} $J_1$ up
to $3 \mu\text{eV}$ ($0.75 \text{GHz}$) has been reported.
In order to initialize the system in its ground-state solely through cooling, the energy gap, $\Delta E$, has to be larger than the temperature of the fridge, typically around $T \sim 50 \text{mK}$ (i.e. $k_B T = 4.3 \mu\text{eV}$ for dilution fridges.
This currently limits direct initialization to short chains ($N\sim 6$).
However, for longer chains, a series of double dot singlets can be adiabatically welded to form the ground-state even in higher temperatures in a time-scale much less than the thermalization time~\cite{farooq_adiabatic_2015}.


We now describe how the triplet profile is measured once the target state is realised in the quantum dot array.
We rapidly (with respect to $1/J_1$) raise voltage barriers to isolate pairs of quantum double dots, each of which can act (when connected to an appropriate circuit) as a capacitor, with capacitance dependent on whether the spin state is a singlet or a triplet.
  It is important that isolating the pairs is rapid in order to avoid any adiabatic evolution towards the new effective Hamiltonian, which would change the state.
When all these capacitors are connected in parallel in a LC-circuit a single reading of the total capacitance measures their sum, from which $m_t$ can be deduced.
Such a circuit is shown in Fig.~\ref{fig:schematic2}(b), which measures the total capacitance of 4 quantum double dots ($N=8$) in a single shot.
The set-up uses dispersive gate sensors coupled to DC gate electrodes via bias tees~\cite{colless2013dispersive}.
The inductors, together with the parasitic capacitance $C_p$, form the resonant circuit with the dot array and thus one can sense the capacitance through the phase and amplitude of the reflected RF signal.

\begin{figure}[]
  \centering
  \includegraphics[width=\linewidth]{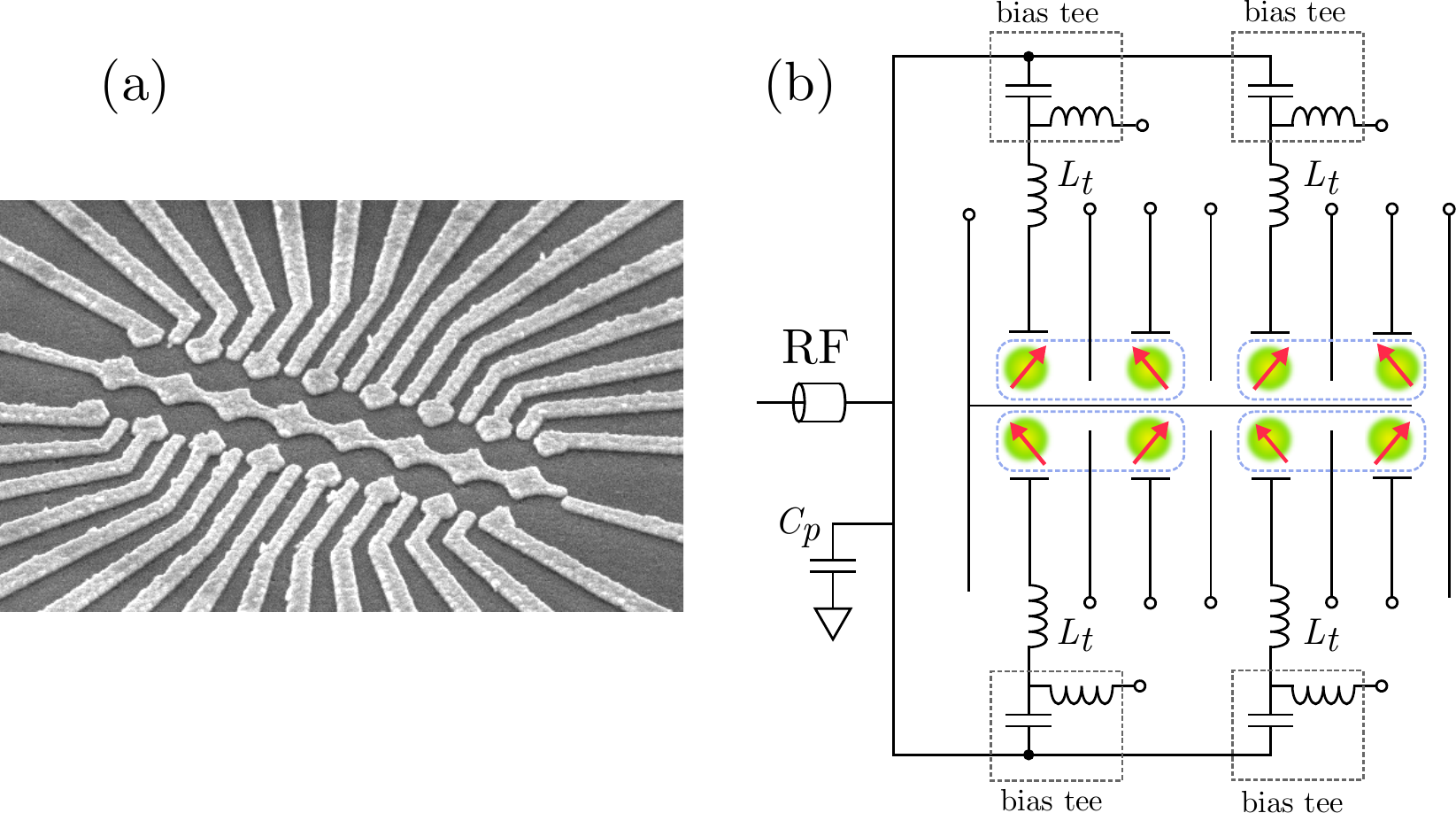}
  \caption{
    (Color online)
  \textbf{A realistic quantum dot array with triplet profile measurement:
  (a)}
  SEM image of an example quantum dot array, as realised in~\cite{puddy_multiplexed_2015}.
  \textbf{(b)}
  Circuit schematic of an eight dot device with dispersive gate sensors which forms a $N=8$ open spin chain.
  The gate sensors are formed by coupling an RF-signal to gate electrodes via bias tees.
  $L_t$ are chosen so that a resonant tank circuit is formed in combination with the dot system and the parasitic capacitance, $C_p$.
  The reflected RF-signal is used to read out the total capacitance of all double dots in parallel and hence $m_t$.
  The dashed blue lines denote pairings of the quantum dots for this ST-measurement.}
  \label{fig:schematic2}
\end{figure}

\section{Conclusions}

Motivated by established technology, we have explored the possibility of using solely singlet-triplet measurements to characterise the achieved ground-states of quantum dot arrays and found that a measurement of a triplet profile is largely sufficient for distinguishing the ground-state from other potential candidates.
Features of this quantity can also indicate that the achieved state is highly non-classical.
Our investigation fits with experimental accessibility as we only demand nearest neighbour measurements, do not demand a full Bell-basis measurement (although this can be achieved in principle with further single qubit rotations), and motivated by scalability do not even demand positional information of the outcomes: only $m_t$, as shown in Fig.~\ref{fig:schematic}.
To demonstrate its utility, we investigated the $J_1-J_2$ phase transition in the Heisenberg ladder.

Since our method is suitable for any models with isotropic antiferromagnetic couplings, one could consider in the future investigating 2D arrays and other more complex geometries. For non-Heisenberg Hamiltonians, such as the Ising model with transverse field, we expect that singlet-triplet measurements are still useful since different phases tend to have different local correlations.
Another clear direction would be to consider the extra information currently missed by only recording the total number of triplets. For example, if information regarding the clustering of the triplet occurrences was retained, this could serve as a second axis on the probability profile. Such an increase in the probability distribution space would clearly aid in distinguishing quantum states, and would also likely reflect physical traits of the system such as correlation length.

As well as characterization, we showed that singlet-triplet measurements have a quantum information processing application in localizing entanglement between the opposite ends of an open SU(2)-symmetric chain.
Engineering the couplings slightly allows this effect to be amplified, though the effect on the Hamiltonian's resultant spectrum must be considered.
Finally, we considered the relevant noise sources for practical application of these techniques in GaAs quantum dot arrays for example.

\emph{Acknowledgements} -
JG acknowledges funding from the EPSRC Center for Doctoral Training in Delivering Quantum Technologies at UCL.
AB, RKP, CGS and SB acknowledge the EPSRC grant EP/K004077/1.
SB acknowledges financial support by the ERC under Starting Grant 308253 PACOMANEDIA.


\begin{thebibliography}{51}
\expandafter\ifx\csname natexlab\endcsname\relax\def\natexlab#1{#1}\fi
\expandafter\ifx\csname bibnamefont\endcsname\relax
  \def\bibnamefont#1{#1}\fi
\expandafter\ifx\csname bibfnamefont\endcsname\relax
  \def\bibfnamefont#1{#1}\fi
\expandafter\ifx\csname citenamefont\endcsname\relax
  \def\citenamefont#1{#1}\fi
\expandafter\ifx\csname url\endcsname\relax
  \def\url#1{\texttt{#1}}\fi
\expandafter\ifx\csname urlprefix\endcsname\relax\def\urlprefix{URL }\fi
\providecommand{\bibinfo}[2]{#2}
\providecommand{\eprint}[2][]{\url{#2}}

\bibitem[{\citenamefont{Buluta and Nori}(2009)}]{buluta2009quantum}
\bibinfo{author}{\bibfnamefont{I.}~\bibnamefont{Buluta}} \bibnamefont{and}
  \bibinfo{author}{\bibfnamefont{F.}~\bibnamefont{Nori}},
  \bibinfo{journal}{Science} \textbf{\bibinfo{volume}{326}},
  \bibinfo{pages}{108} (\bibinfo{year}{2009}).

\bibitem[{\citenamefont{Stafford and {Das
  Sarma}}(1994)}]{stafford_collective_1994}
\bibinfo{author}{\bibfnamefont{C.~A.} \bibnamefont{Stafford}} \bibnamefont{and}
  \bibinfo{author}{\bibfnamefont{S.}~\bibnamefont{{Das Sarma}}},
  \bibinfo{journal}{Phys. Rev. Lett.} \textbf{\bibinfo{volume}{72}},
  \bibinfo{pages}{3590} (\bibinfo{year}{1994}).

\bibitem[{\citenamefont{Barthelemy and
  Vandersypen}(2013)}]{barthelemy2013quantum}
\bibinfo{author}{\bibfnamefont{P.}~\bibnamefont{Barthelemy}} \bibnamefont{and}
  \bibinfo{author}{\bibfnamefont{L.~M.} \bibnamefont{Vandersypen}},
  \bibinfo{journal}{Annalen der Physik} \textbf{\bibinfo{volume}{525}},
  \bibinfo{pages}{808} (\bibinfo{year}{2013}).

\bibitem[{\citenamefont{Yang et~al.}(2010)\citenamefont{Yang, Bayat, and
  Bose}}]{yang_spin-state_2010}
\bibinfo{author}{\bibfnamefont{S.}~\bibnamefont{Yang}},
  \bibinfo{author}{\bibfnamefont{A.}~\bibnamefont{Bayat}}, \bibnamefont{and}
  \bibinfo{author}{\bibfnamefont{S.}~\bibnamefont{Bose}},
  \bibinfo{journal}{Phys. Rev A} \textbf{\bibinfo{volume}{82}},
  \bibinfo{pages}{022336} (\bibinfo{year}{2010}).

\bibitem[{\citenamefont{Farooq et~al.}(2015)\citenamefont{Farooq, Bayat,
  Mancini, and Bose}}]{farooq_adiabatic_2015}
\bibinfo{author}{\bibfnamefont{U.}~\bibnamefont{Farooq}},
  \bibinfo{author}{\bibfnamefont{A.}~\bibnamefont{Bayat}},
  \bibinfo{author}{\bibfnamefont{S.}~\bibnamefont{Mancini}}, \bibnamefont{and}
  \bibinfo{author}{\bibfnamefont{S.}~\bibnamefont{Bose}},
  \bibinfo{journal}{Phys. Rev. B} \textbf{\bibinfo{volume}{91}},
  \bibinfo{pages}{134303} (\bibinfo{year}{2015}).

\bibitem[{\citenamefont{Salfi et~al.}(2016)\citenamefont{Salfi, Mol, Rahman,
  Klimeck, Simmons, Hollenberg, and Rogge}}]{salfi_quantum_2016}
\bibinfo{author}{\bibfnamefont{J.}~\bibnamefont{Salfi}},
  \bibinfo{author}{\bibfnamefont{J.~A.} \bibnamefont{Mol}},
  \bibinfo{author}{\bibfnamefont{R.}~\bibnamefont{Rahman}},
  \bibinfo{author}{\bibfnamefont{G.}~\bibnamefont{Klimeck}},
  \bibinfo{author}{\bibfnamefont{M.~Y.} \bibnamefont{Simmons}},
  \bibinfo{author}{\bibfnamefont{L.~C.~L.} \bibnamefont{Hollenberg}},
  \bibnamefont{and} \bibinfo{author}{\bibfnamefont{S.}~\bibnamefont{Rogge}},
  \bibinfo{journal}{Nat. Commun.} \textbf{\bibinfo{volume}{7}},
  \bibinfo{pages}{11342} (\bibinfo{year}{2016}).

\bibitem[{\citenamefont{Prati et~al.}(2012)\citenamefont{Prati, Hori,
  Guagliardo, Ferrari, and Shinada}}]{prati2012anderson}
\bibinfo{author}{\bibfnamefont{E.}~\bibnamefont{Prati}},
  \bibinfo{author}{\bibfnamefont{M.}~\bibnamefont{Hori}},
  \bibinfo{author}{\bibfnamefont{F.}~\bibnamefont{Guagliardo}},
  \bibinfo{author}{\bibfnamefont{G.}~\bibnamefont{Ferrari}}, \bibnamefont{and}
  \bibinfo{author}{\bibfnamefont{T.}~\bibnamefont{Shinada}},
  \bibinfo{journal}{Nat. Nano.} \textbf{\bibinfo{volume}{7}},
  \bibinfo{pages}{443} (\bibinfo{year}{2012}).

\bibitem[{\citenamefont{Bloch et~al.}(2012)\citenamefont{Bloch, Dalibard, and
  Nascimb{\`e}ne}}]{bloch_quantum_2012}
\bibinfo{author}{\bibfnamefont{I.}~\bibnamefont{Bloch}},
  \bibinfo{author}{\bibfnamefont{J.}~\bibnamefont{Dalibard}}, \bibnamefont{and}
  \bibinfo{author}{\bibfnamefont{S.}~\bibnamefont{Nascimb{\`e}ne}},
  \bibinfo{journal}{Nat. Phys.} \textbf{\bibinfo{volume}{8}},
  \bibinfo{pages}{267} (\bibinfo{year}{2012}), ISSN \bibinfo{issn}{1745-2473}.

\bibitem[{\citenamefont{Blatt and Roos}(2012)}]{blatt_quantum_2012}
\bibinfo{author}{\bibfnamefont{R.}~\bibnamefont{Blatt}} \bibnamefont{and}
  \bibinfo{author}{\bibfnamefont{C.~F.} \bibnamefont{Roos}},
  \bibinfo{journal}{Nat. Phys.} \textbf{\bibinfo{volume}{8}},
  \bibinfo{pages}{277} (\bibinfo{year}{2012}), ISSN \bibinfo{issn}{1745-2473}.

\bibitem[{\citenamefont{Puddy et~al.}(2015)\citenamefont{Puddy, Smith, Al-Taie,
  Chong, Farrer, Griffiths, Ritchie, Kelly, Pepper, and
  Smith}}]{puddy_multiplexed_2015}
\bibinfo{author}{\bibfnamefont{R.~K.} \bibnamefont{Puddy}},
  \bibinfo{author}{\bibfnamefont{L.~W.} \bibnamefont{Smith}},
  \bibinfo{author}{\bibfnamefont{H.}~\bibnamefont{Al-Taie}},
  \bibinfo{author}{\bibfnamefont{C.~H.} \bibnamefont{Chong}},
  \bibinfo{author}{\bibfnamefont{I.}~\bibnamefont{Farrer}},
  \bibinfo{author}{\bibfnamefont{J.~P.} \bibnamefont{Griffiths}},
  \bibinfo{author}{\bibfnamefont{D.~A.} \bibnamefont{Ritchie}},
  \bibinfo{author}{\bibfnamefont{M.~J.} \bibnamefont{Kelly}},
  \bibinfo{author}{\bibfnamefont{M.}~\bibnamefont{Pepper}}, \bibnamefont{and}
  \bibinfo{author}{\bibfnamefont{C.~G.} \bibnamefont{Smith}},
  \bibinfo{journal}{Appl. Phys. Lett.} \textbf{\bibinfo{volume}{107}},
  \bibinfo{pages}{143501} (\bibinfo{year}{2015}), ISSN
  \bibinfo{issn}{0003-6951, 1077-3118}.

\bibitem[{\citenamefont{Loss and DiVincenzo}(1998)}]{loss_quantum_1998}
\bibinfo{author}{\bibfnamefont{D.}~\bibnamefont{Loss}} \bibnamefont{and}
  \bibinfo{author}{\bibfnamefont{D.~P.} \bibnamefont{DiVincenzo}},
  \bibinfo{journal}{Phys. Rev A} \textbf{\bibinfo{volume}{57}},
  \bibinfo{pages}{120} (\bibinfo{year}{1998}).

\bibitem[{\citenamefont{Hanson et~al.}(2007)\citenamefont{Hanson, Kouwenhoven,
  Petta, Tarucha, and Vandersypen}}]{hanson2007spins}
\bibinfo{author}{\bibfnamefont{R.}~\bibnamefont{Hanson}},
  \bibinfo{author}{\bibfnamefont{L.}~\bibnamefont{Kouwenhoven}},
  \bibinfo{author}{\bibfnamefont{J.}~\bibnamefont{Petta}},
  \bibinfo{author}{\bibfnamefont{S.}~\bibnamefont{Tarucha}}, \bibnamefont{and}
  \bibinfo{author}{\bibfnamefont{L.}~\bibnamefont{Vandersypen}},
  \bibinfo{journal}{Rev. Mod. Phys} \textbf{\bibinfo{volume}{79}},
  \bibinfo{pages}{1217} (\bibinfo{year}{2007}).

\bibitem[{\citenamefont{Petta et~al.}(2005)\citenamefont{Petta, Johnson,
  Taylor, Laird, Yacoby, Lukin, Marcus, Hanson, and
  Gossard}}]{petta_coherent_2005}
\bibinfo{author}{\bibfnamefont{J.~R.} \bibnamefont{Petta}},
  \bibinfo{author}{\bibfnamefont{A.~C.} \bibnamefont{Johnson}},
  \bibinfo{author}{\bibfnamefont{J.~M.} \bibnamefont{Taylor}},
  \bibinfo{author}{\bibfnamefont{E.~A.} \bibnamefont{Laird}},
  \bibinfo{author}{\bibfnamefont{A.}~\bibnamefont{Yacoby}},
  \bibinfo{author}{\bibfnamefont{M.~D.} \bibnamefont{Lukin}},
  \bibinfo{author}{\bibfnamefont{C.~M.} \bibnamefont{Marcus}},
  \bibinfo{author}{\bibfnamefont{M.~P.} \bibnamefont{Hanson}},
  \bibnamefont{and} \bibinfo{author}{\bibfnamefont{A.~C.}
  \bibnamefont{Gossard}}, \bibinfo{journal}{Science}
  \textbf{\bibinfo{volume}{309}}, \bibinfo{pages}{2180} (\bibinfo{year}{2005}),
  ISSN \bibinfo{issn}{0036-8075, 1095-9203}.

\bibitem[{\citenamefont{Levy}(2002)}]{levy_universal_2002}
\bibinfo{author}{\bibfnamefont{J.}~\bibnamefont{Levy}}, \bibinfo{journal}{Phys.
  Rev. Lett.} \textbf{\bibinfo{volume}{89}}, \bibinfo{pages}{147902}
  (\bibinfo{year}{2002}).

\bibitem[{\citenamefont{Petersson et~al.}(2010)\citenamefont{Petersson, Smith,
  Anderson, Atkinson, Jones, and Ritchie}}]{petersson2010charge}
\bibinfo{author}{\bibfnamefont{K.}~\bibnamefont{Petersson}},
  \bibinfo{author}{\bibfnamefont{C.}~\bibnamefont{Smith}},
  \bibinfo{author}{\bibfnamefont{D.}~\bibnamefont{Anderson}},
  \bibinfo{author}{\bibfnamefont{P.}~\bibnamefont{Atkinson}},
  \bibinfo{author}{\bibfnamefont{G.}~\bibnamefont{Jones}}, \bibnamefont{and}
  \bibinfo{author}{\bibfnamefont{D.}~\bibnamefont{Ritchie}},
  \bibinfo{journal}{Nano Lett.} \textbf{\bibinfo{volume}{10}},
  \bibinfo{pages}{2789} (\bibinfo{year}{2010}).

\bibitem[{\citenamefont{Jung et~al.}(2012)\citenamefont{Jung, Schroer,
  Petersson, and Petta}}]{jung2012radio}
\bibinfo{author}{\bibfnamefont{M.}~\bibnamefont{Jung}},
  \bibinfo{author}{\bibfnamefont{M.}~\bibnamefont{Schroer}},
  \bibinfo{author}{\bibfnamefont{K.}~\bibnamefont{Petersson}},
  \bibnamefont{and} \bibinfo{author}{\bibfnamefont{J.}~\bibnamefont{Petta}},
  \bibinfo{journal}{App. Phys. Lett.} \textbf{\bibinfo{volume}{100}},
  \bibinfo{pages}{253508} (\bibinfo{year}{2012}).

\bibitem[{\citenamefont{Basset et~al.}(2013)\citenamefont{Basset, Jarausch,
  Stockklauser, Frey, Reichl, Wegscheider, Ihn, Ensslin, and
  Wallraff}}]{basset2013single}
\bibinfo{author}{\bibfnamefont{J.}~\bibnamefont{Basset}},
  \bibinfo{author}{\bibfnamefont{D.-D.} \bibnamefont{Jarausch}},
  \bibinfo{author}{\bibfnamefont{A.}~\bibnamefont{Stockklauser}},
  \bibinfo{author}{\bibfnamefont{T.}~\bibnamefont{Frey}},
  \bibinfo{author}{\bibfnamefont{C.}~\bibnamefont{Reichl}},
  \bibinfo{author}{\bibfnamefont{W.}~\bibnamefont{Wegscheider}},
  \bibinfo{author}{\bibfnamefont{T.~M.} \bibnamefont{Ihn}},
  \bibinfo{author}{\bibfnamefont{K.}~\bibnamefont{Ensslin}}, \bibnamefont{and}
  \bibinfo{author}{\bibfnamefont{A.}~\bibnamefont{Wallraff}},
  \bibinfo{journal}{Phys. Rev. B} \textbf{\bibinfo{volume}{88}},
  \bibinfo{pages}{125312} (\bibinfo{year}{2013}).

\bibitem[{\citenamefont{Chorley et~al.}(2012)\citenamefont{Chorley, Wabnig,
  Penfold-Fitch, Petersson, Frake, Smith, and
  Buitelaar}}]{chorley2012measuring}
\bibinfo{author}{\bibfnamefont{S.}~\bibnamefont{Chorley}},
  \bibinfo{author}{\bibfnamefont{J.}~\bibnamefont{Wabnig}},
  \bibinfo{author}{\bibfnamefont{Z.}~\bibnamefont{Penfold-Fitch}},
  \bibinfo{author}{\bibfnamefont{K.}~\bibnamefont{Petersson}},
  \bibinfo{author}{\bibfnamefont{J.}~\bibnamefont{Frake}},
  \bibinfo{author}{\bibfnamefont{C.}~\bibnamefont{Smith}}, \bibnamefont{and}
  \bibinfo{author}{\bibfnamefont{M.}~\bibnamefont{Buitelaar}},
  \bibinfo{journal}{Phys. Rev. Lett.} \textbf{\bibinfo{volume}{108}},
  \bibinfo{pages}{036802} (\bibinfo{year}{2012}).

\bibitem[{\citenamefont{Frey et~al.}(2012)\citenamefont{Frey, Leek, Beck,
  Blais, Ihn, Ensslin, and Wallraff}}]{frey2012dipole}
\bibinfo{author}{\bibfnamefont{T.}~\bibnamefont{Frey}},
  \bibinfo{author}{\bibfnamefont{P.}~\bibnamefont{Leek}},
  \bibinfo{author}{\bibfnamefont{M.}~\bibnamefont{Beck}},
  \bibinfo{author}{\bibfnamefont{A.}~\bibnamefont{Blais}},
  \bibinfo{author}{\bibfnamefont{T.}~\bibnamefont{Ihn}},
  \bibinfo{author}{\bibfnamefont{K.}~\bibnamefont{Ensslin}}, \bibnamefont{and}
  \bibinfo{author}{\bibfnamefont{A.}~\bibnamefont{Wallraff}},
  \bibinfo{journal}{Phys. Rev. Lett.} \textbf{\bibinfo{volume}{108}},
  \bibinfo{pages}{046807} (\bibinfo{year}{2012}).

\bibitem[{\citenamefont{Colless et~al.}(2013)\citenamefont{Colless, Mahoney,
  Hornibrook, Doherty, Lu, Gossard, and Reilly}}]{colless2013dispersive}
\bibinfo{author}{\bibfnamefont{J.}~\bibnamefont{Colless}},
  \bibinfo{author}{\bibfnamefont{A.}~\bibnamefont{Mahoney}},
  \bibinfo{author}{\bibfnamefont{J.}~\bibnamefont{Hornibrook}},
  \bibinfo{author}{\bibfnamefont{A.}~\bibnamefont{Doherty}},
  \bibinfo{author}{\bibfnamefont{H.}~\bibnamefont{Lu}},
  \bibinfo{author}{\bibfnamefont{A.}~\bibnamefont{Gossard}}, \bibnamefont{and}
  \bibinfo{author}{\bibfnamefont{D.}~\bibnamefont{Reilly}},
  \bibinfo{journal}{Phys. Rev. Lett,} \textbf{\bibinfo{volume}{110}},
  \bibinfo{pages}{046805} (\bibinfo{year}{2013}).

\bibitem[{\citenamefont{House et~al.}(2015)\citenamefont{House, Kobayashi,
  Weber, Hile, Watson, van~der Heijden, Rogge, and Simmons}}]{house2015radio}
\bibinfo{author}{\bibfnamefont{M.}~\bibnamefont{House}},
  \bibinfo{author}{\bibfnamefont{T.}~\bibnamefont{Kobayashi}},
  \bibinfo{author}{\bibfnamefont{B.}~\bibnamefont{Weber}},
  \bibinfo{author}{\bibfnamefont{S.}~\bibnamefont{Hile}},
  \bibinfo{author}{\bibfnamefont{T.}~\bibnamefont{Watson}},
  \bibinfo{author}{\bibfnamefont{J.}~\bibnamefont{van~der Heijden}},
  \bibinfo{author}{\bibfnamefont{S.}~\bibnamefont{Rogge}}, \bibnamefont{and}
  \bibinfo{author}{\bibfnamefont{M.}~\bibnamefont{Simmons}},
  \bibinfo{journal}{Nat. Comm.} \textbf{\bibinfo{volume}{6}}
  (\bibinfo{year}{2015}).

\bibitem[{\citenamefont{Rudolph and Virmani}(2005)}]{rudolph_relational_2005}
\bibinfo{author}{\bibfnamefont{T.}~\bibnamefont{Rudolph}} \bibnamefont{and}
  \bibinfo{author}{\bibfnamefont{S.~S.} \bibnamefont{Virmani}},
  \bibinfo{journal}{New J. Phys} \textbf{\bibinfo{volume}{7}},
  \bibinfo{pages}{228} (\bibinfo{year}{2005}), ISSN \bibinfo{issn}{1367-2630}.

\bibitem[{\citenamefont{D'Ariano et~al.}(2003)\citenamefont{D'Ariano, Paris,
  and Sacchi}}]{quantum_tomography_book}
\bibinfo{author}{\bibfnamefont{G.~M.} \bibnamefont{D'Ariano}},
  \bibinfo{author}{\bibfnamefont{M.~G.} \bibnamefont{Paris}}, \bibnamefont{and}
  \bibinfo{author}{\bibfnamefont{M.~F.} \bibnamefont{Sacchi}},
  \bibinfo{journal}{Advances in Imaging and Electron Physics}
  \textbf{\bibinfo{volume}{128}}, \bibinfo{pages}{206} (\bibinfo{year}{2003}).

\bibitem[{\citenamefont{Gross et~al.}(2010)\citenamefont{Gross, Liu, Flammia,
  Becker, and Eisert}}]{gross_quantum_2010}
\bibinfo{author}{\bibfnamefont{D.}~\bibnamefont{Gross}},
  \bibinfo{author}{\bibfnamefont{Y.-K.} \bibnamefont{Liu}},
  \bibinfo{author}{\bibfnamefont{S.~T.} \bibnamefont{Flammia}},
  \bibinfo{author}{\bibfnamefont{S.}~\bibnamefont{Becker}}, \bibnamefont{and}
  \bibinfo{author}{\bibfnamefont{J.}~\bibnamefont{Eisert}},
  \bibinfo{journal}{Phys. Rev. Lett.} \textbf{\bibinfo{volume}{105}}
  (\bibinfo{year}{2010}), ISSN \bibinfo{issn}{0031-9007, 1079-7114},
  \eprint{0909.3304}.

\bibitem[{\citenamefont{Cramer et~al.}(2010)\citenamefont{Cramer, Plenio,
  Flammia, Somma, Gross, Bartlett, Landon-Cardinal, Poulin, and
  Liu}}]{cramer_efficient_2010-1}
\bibinfo{author}{\bibfnamefont{M.}~\bibnamefont{Cramer}},
  \bibinfo{author}{\bibfnamefont{M.~B.} \bibnamefont{Plenio}},
  \bibinfo{author}{\bibfnamefont{S.~T.} \bibnamefont{Flammia}},
  \bibinfo{author}{\bibfnamefont{R.}~\bibnamefont{Somma}},
  \bibinfo{author}{\bibfnamefont{D.}~\bibnamefont{Gross}},
  \bibinfo{author}{\bibfnamefont{S.~D.} \bibnamefont{Bartlett}},
  \bibinfo{author}{\bibfnamefont{O.}~\bibnamefont{Landon-Cardinal}},
  \bibinfo{author}{\bibfnamefont{D.}~\bibnamefont{Poulin}}, \bibnamefont{and}
  \bibinfo{author}{\bibfnamefont{Y.-K.} \bibnamefont{Liu}},
  \bibinfo{journal}{Nat. Commun.} \textbf{\bibinfo{volume}{1}},
  \bibinfo{pages}{149} (\bibinfo{year}{2010}).

\bibitem[{\citenamefont{Barthel et~al.}(2009)\citenamefont{Barthel, Reilly,
  Marcus, Hanson, and Gossard}}]{barthel_rapid_2009}
\bibinfo{author}{\bibfnamefont{C.}~\bibnamefont{Barthel}},
  \bibinfo{author}{\bibfnamefont{D.~J.} \bibnamefont{Reilly}},
  \bibinfo{author}{\bibfnamefont{C.~M.} \bibnamefont{Marcus}},
  \bibinfo{author}{\bibfnamefont{M.~P.} \bibnamefont{Hanson}},
  \bibnamefont{and} \bibinfo{author}{\bibfnamefont{A.~C.}
  \bibnamefont{Gossard}}, \bibinfo{journal}{Phys. Rev. Lett.}
  \textbf{\bibinfo{volume}{103}}, \bibinfo{pages}{160503}
  (\bibinfo{year}{2009}).

\bibitem[{\citenamefont{Shulman et~al.}(2012)\citenamefont{Shulman, Dial,
  Harvey, Bluhm, Umansky, and Yacoby}}]{shulman_demonstration_2012}
\bibinfo{author}{\bibfnamefont{M.~D.} \bibnamefont{Shulman}},
  \bibinfo{author}{\bibfnamefont{O.~E.} \bibnamefont{Dial}},
  \bibinfo{author}{\bibfnamefont{S.~P.} \bibnamefont{Harvey}},
  \bibinfo{author}{\bibfnamefont{H.}~\bibnamefont{Bluhm}},
  \bibinfo{author}{\bibfnamefont{V.}~\bibnamefont{Umansky}}, \bibnamefont{and}
  \bibinfo{author}{\bibfnamefont{A.}~\bibnamefont{Yacoby}},
  \bibinfo{journal}{Science} \textbf{\bibinfo{volume}{336}},
  \bibinfo{pages}{202} (\bibinfo{year}{2012}), ISSN \bibinfo{issn}{0036-8075,
  1095-9203}.

\bibitem[{\citenamefont{Blundell}(2001)}]{blundell2001magnetism}
\bibinfo{author}{\bibfnamefont{S.}~\bibnamefont{Blundell}},
  \emph{\bibinfo{title}{Magnetism in condensed matter}}
  (\bibinfo{publisher}{Oxford University Press}, \bibinfo{year}{2001}).

\bibitem[{\citenamefont{Sachdev}(2011)}]{sachdev2011quantum}
\bibinfo{author}{\bibfnamefont{S.}~\bibnamefont{Sachdev}},
  \emph{\bibinfo{title}{Quantum phase transitions}}
  (\bibinfo{publisher}{Cambridge University Press}, \bibinfo{year}{2011}).

\bibitem[{\citenamefont{Bose}(2003)}]{bose2003quantum}
\bibinfo{author}{\bibfnamefont{S.}~\bibnamefont{Bose}}, \bibinfo{journal}{Phys.
  Rev. Lett.} \textbf{\bibinfo{volume}{91}}, \bibinfo{pages}{207901}
  (\bibinfo{year}{2003}).

\bibitem[{\citenamefont{Nikolopoulos and Jex}(2014)}]{nikolopoulos2014quantum}
\bibinfo{author}{\bibfnamefont{G.~M.} \bibnamefont{Nikolopoulos}}
  \bibnamefont{and} \bibinfo{author}{\bibfnamefont{I.}~\bibnamefont{Jex}},
  \emph{\bibinfo{title}{Quantum State Transfer and Network Engineering}}
  (\bibinfo{publisher}{Springer}, \bibinfo{year}{2014}).

\bibitem[{\citenamefont{Nielsen and Chuang}(2010)}]{nielsen2010quantum}
\bibinfo{author}{\bibfnamefont{M.~A.} \bibnamefont{Nielsen}} \bibnamefont{and}
  \bibinfo{author}{\bibfnamefont{I.~L.} \bibnamefont{Chuang}},
  \emph{\bibinfo{title}{Quantum computation and quantum information}}
  (\bibinfo{publisher}{Cambridge university press}, \bibinfo{year}{2010}).

\bibitem[{\citenamefont{O'Connor and Wootters}(2001)}]{oconnor_entangled_2001}
\bibinfo{author}{\bibfnamefont{K.~M.} \bibnamefont{O'Connor}} \bibnamefont{and}
  \bibinfo{author}{\bibfnamefont{W.~K.} \bibnamefont{Wootters}},
  \bibinfo{journal}{Phys. Rev A} \textbf{\bibinfo{volume}{63}},
  \bibinfo{pages}{052302} (\bibinfo{year}{2001}).

\bibitem[{\citenamefont{Baigneres et~al.}(2004)\citenamefont{Baigneres, Junod,
  and Vaudenay}}]{baigneres2004far}
\bibinfo{author}{\bibfnamefont{T.}~\bibnamefont{Baigneres}},
  \bibinfo{author}{\bibfnamefont{P.}~\bibnamefont{Junod}}, \bibnamefont{and}
  \bibinfo{author}{\bibfnamefont{S.}~\bibnamefont{Vaudenay}}, in
  \emph{\bibinfo{booktitle}{Advances in Cryptology-Asiacrypt 2004}}
  (\bibinfo{publisher}{Springer}, \bibinfo{year}{2004}), pp.
  \bibinfo{pages}{432--450}.

\bibitem[{\citenamefont{Helstrom}(1969)}]{helstrom_quantum_1969}
\bibinfo{author}{\bibfnamefont{C.~W.} \bibnamefont{Helstrom}},
  \bibinfo{journal}{J. Stat. Phys.} \textbf{\bibinfo{volume}{1}},
  \bibinfo{pages}{231} (\bibinfo{year}{1969}), ISSN \bibinfo{issn}{0022-4715,
  1572-9613}.

\bibitem[{\citenamefont{Thesberg and S{\o}rensen}(2011)}]{thesberg2011general}
\bibinfo{author}{\bibfnamefont{M.}~\bibnamefont{Thesberg}} \bibnamefont{and}
  \bibinfo{author}{\bibfnamefont{E.~S.} \bibnamefont{S{\o}rensen}},
  \bibinfo{journal}{Phys. Rev. B} \textbf{\bibinfo{volume}{84}},
  \bibinfo{pages}{224435} (\bibinfo{year}{2011}).

\bibitem[{\citenamefont{Chen et~al.}(2007)\citenamefont{Chen, Wang, Gu, and
  Wang}}]{chen_fidelity_2007}
\bibinfo{author}{\bibfnamefont{S.}~\bibnamefont{Chen}},
  \bibinfo{author}{\bibfnamefont{L.}~\bibnamefont{Wang}},
  \bibinfo{author}{\bibfnamefont{S.-J.} \bibnamefont{Gu}}, \bibnamefont{and}
  \bibinfo{author}{\bibfnamefont{Y.}~\bibnamefont{Wang}},
  \bibinfo{journal}{Phys. Rev. E} \textbf{\bibinfo{volume}{76}},
  \bibinfo{pages}{061108} (\bibinfo{year}{2007}).

\bibitem[{\citenamefont{Verstraete et~al.}(2004)\citenamefont{Verstraete, Popp,
  and Cirac}}]{verstraete_entanglement_2004}
\bibinfo{author}{\bibfnamefont{F.}~\bibnamefont{Verstraete}},
  \bibinfo{author}{\bibfnamefont{M.}~\bibnamefont{Popp}}, \bibnamefont{and}
  \bibinfo{author}{\bibfnamefont{J.~I.} \bibnamefont{Cirac}},
  \bibinfo{journal}{Phys. Rev. Lett.} \textbf{\bibinfo{volume}{92}},
  \bibinfo{pages}{027901} (\bibinfo{year}{2004}).

\bibitem[{\citenamefont{Popp et~al.}(2005)\citenamefont{Popp, Verstraete,
  Mart{\'\i}n-Delgado, and Cirac}}]{popp_localizable_2005}
\bibinfo{author}{\bibfnamefont{M.}~\bibnamefont{Popp}},
  \bibinfo{author}{\bibfnamefont{F.}~\bibnamefont{Verstraete}},
  \bibinfo{author}{\bibfnamefont{M.~A.} \bibnamefont{Mart{\'\i}n-Delgado}},
  \bibnamefont{and} \bibinfo{author}{\bibfnamefont{J.~I.} \bibnamefont{Cirac}},
  \bibinfo{journal}{Phys. Rev. A} \textbf{\bibinfo{volume}{71}},
  \bibinfo{pages}{042306} (\bibinfo{year}{2005}).

\bibitem[{\citenamefont{{Campos Venuti} et~al.}(2007)\citenamefont{{Campos
  Venuti}, {Degli Esposti Boschi}, and Roncaglia}}]{campos_venuti_qubit_2007}
\bibinfo{author}{\bibfnamefont{L.}~\bibnamefont{{Campos Venuti}}},
  \bibinfo{author}{\bibfnamefont{C.}~\bibnamefont{{Degli Esposti Boschi}}},
  \bibnamefont{and}
  \bibinfo{author}{\bibfnamefont{M.}~\bibnamefont{Roncaglia}},
  \bibinfo{journal}{Phys. Rev. Lett.} \textbf{\bibinfo{volume}{99}},
  \bibinfo{pages}{060401} (\bibinfo{year}{2007}).

\bibitem[{\citenamefont{Barjaktarevic et~al.}(2005)\citenamefont{Barjaktarevic,
  McKenzie, Links, and Milburn}}]{barjaktarevic_measurement-based_2005}
\bibinfo{author}{\bibfnamefont{J.~P.} \bibnamefont{Barjaktarevic}},
  \bibinfo{author}{\bibfnamefont{R.~H.} \bibnamefont{McKenzie}},
  \bibinfo{author}{\bibfnamefont{J.}~\bibnamefont{Links}}, \bibnamefont{and}
  \bibinfo{author}{\bibfnamefont{G.~J.} \bibnamefont{Milburn}},
  \bibinfo{journal}{Phys. Rev. Lett.} \textbf{\bibinfo{volume}{95}},
  \bibinfo{pages}{230501} (\bibinfo{year}{2005}).

\bibitem[{\citenamefont{Taylor et~al.}(2007)\citenamefont{Taylor, Petta,
  Johnson, Yacoby, Marcus, and Lukin}}]{taylor_relaxation_2007}
\bibinfo{author}{\bibfnamefont{J.~M.} \bibnamefont{Taylor}},
  \bibinfo{author}{\bibfnamefont{J.~R.} \bibnamefont{Petta}},
  \bibinfo{author}{\bibfnamefont{A.~C.} \bibnamefont{Johnson}},
  \bibinfo{author}{\bibfnamefont{A.}~\bibnamefont{Yacoby}},
  \bibinfo{author}{\bibfnamefont{C.~M.} \bibnamefont{Marcus}},
  \bibnamefont{and} \bibinfo{author}{\bibfnamefont{M.~D.} \bibnamefont{Lukin}},
  \bibinfo{journal}{Phys. Rev. B} \textbf{\bibinfo{volume}{76}},
  \bibinfo{pages}{035315} (\bibinfo{year}{2007}).

\bibitem[{\citenamefont{Malinowski et~al.}(2016)\citenamefont{Malinowski,
  Martins, Nissen, Barnes, Rudner, Fallahi, Gardner, Manfra, Marcus, and
  Kuemmeth}}]{malinowski2016notch}
\bibinfo{author}{\bibfnamefont{F.~K.} \bibnamefont{Malinowski}},
  \bibinfo{author}{\bibfnamefont{F.}~\bibnamefont{Martins}},
  \bibinfo{author}{\bibfnamefont{P.~D.} \bibnamefont{Nissen}},
  \bibinfo{author}{\bibfnamefont{E.}~\bibnamefont{Barnes}},
  \bibinfo{author}{\bibfnamefont{M.~S.} \bibnamefont{Rudner}},
  \bibinfo{author}{\bibfnamefont{S.}~\bibnamefont{Fallahi}},
  \bibinfo{author}{\bibfnamefont{G.~C.} \bibnamefont{Gardner}},
  \bibinfo{author}{\bibfnamefont{M.~J.} \bibnamefont{Manfra}},
  \bibinfo{author}{\bibfnamefont{C.~M.} \bibnamefont{Marcus}},
  \bibnamefont{and} \bibinfo{author}{\bibfnamefont{F.}~\bibnamefont{Kuemmeth}},
  \bibinfo{journal}{arXiv:1601.06677 [Nat. Nanotechnol. (to be published)]}
  (\bibinfo{year}{2016}).

\bibitem[{\citenamefont{Kawakami et~al.}(2014)\citenamefont{Kawakami, Scarlino,
  Ward, Braakman, Savage, Lagally, Friesen, Coppersmith, Eriksson, and
  Vandersypen}}]{kawakami2014electrical}
\bibinfo{author}{\bibfnamefont{E.}~\bibnamefont{Kawakami}},
  \bibinfo{author}{\bibfnamefont{P.}~\bibnamefont{Scarlino}},
  \bibinfo{author}{\bibfnamefont{D.}~\bibnamefont{Ward}},
  \bibinfo{author}{\bibfnamefont{F.}~\bibnamefont{Braakman}},
  \bibinfo{author}{\bibfnamefont{D.}~\bibnamefont{Savage}},
  \bibinfo{author}{\bibfnamefont{M.}~\bibnamefont{Lagally}},
  \bibinfo{author}{\bibfnamefont{M.}~\bibnamefont{Friesen}},
  \bibinfo{author}{\bibfnamefont{S.}~\bibnamefont{Coppersmith}},
  \bibinfo{author}{\bibfnamefont{M.}~\bibnamefont{Eriksson}}, \bibnamefont{and}
  \bibinfo{author}{\bibfnamefont{L.}~\bibnamefont{Vandersypen}},
  \bibinfo{journal}{Nat. Nanotechnol.} \textbf{\bibinfo{volume}{9}},
  \bibinfo{pages}{666} (\bibinfo{year}{2014}).

\bibitem[{\citenamefont{Hill and Wootters}(1997)}]{hill_entanglement_1997}
\bibinfo{author}{\bibfnamefont{S.}~\bibnamefont{Hill}} \bibnamefont{and}
  \bibinfo{author}{\bibfnamefont{W.~K.} \bibnamefont{Wootters}},
  \bibinfo{journal}{Phys. Rev. Lett.} \textbf{\bibinfo{volume}{78}},
  \bibinfo{pages}{5022} (\bibinfo{year}{1997}).

\bibitem[{\citenamefont{Wardrop and
  Doherty}(2016)}]{wardrop2016characterization}
\bibinfo{author}{\bibfnamefont{M.~P.} \bibnamefont{Wardrop}} \bibnamefont{and}
  \bibinfo{author}{\bibfnamefont{A.~C.} \bibnamefont{Doherty}},
  \bibinfo{journal}{Phys. Rev. B} \textbf{\bibinfo{volume}{93}},
  \bibinfo{pages}{075436} (\bibinfo{year}{2016}).

\bibitem[{\citenamefont{Petrosyan and
  Lambropoulos}(2006)}]{petrosyan2006coherent}
\bibinfo{author}{\bibfnamefont{D.}~\bibnamefont{Petrosyan}} \bibnamefont{and}
  \bibinfo{author}{\bibfnamefont{P.}~\bibnamefont{Lambropoulos}},
  \bibinfo{journal}{Opt. Commun} \textbf{\bibinfo{volume}{264}},
  \bibinfo{pages}{419} (\bibinfo{year}{2006}).

\bibitem[{\citenamefont{Petrosyan et~al.}(2010)\citenamefont{Petrosyan,
  Nikolopoulos, and Lambropoulos}}]{petrosyan2010state}
\bibinfo{author}{\bibfnamefont{D.}~\bibnamefont{Petrosyan}},
  \bibinfo{author}{\bibfnamefont{G.~M.} \bibnamefont{Nikolopoulos}},
  \bibnamefont{and}
  \bibinfo{author}{\bibfnamefont{P.}~\bibnamefont{Lambropoulos}},
  \bibinfo{journal}{Phys. Rev. A} \textbf{\bibinfo{volume}{81}},
  \bibinfo{pages}{042307} (\bibinfo{year}{2010}).

\bibitem[{\citenamefont{Ares et~al.}(2016)\citenamefont{Ares, Schupp,
  Mavalankar, Rogers, Griffiths, Jones, Farrer, Ritchie, Smith, Cottet
  et~al.}}]{ares2016sensitive}
\bibinfo{author}{\bibfnamefont{N.}~\bibnamefont{Ares}},
  \bibinfo{author}{\bibfnamefont{F.}~\bibnamefont{Schupp}},
  \bibinfo{author}{\bibfnamefont{A.}~\bibnamefont{Mavalankar}},
  \bibinfo{author}{\bibfnamefont{G.}~\bibnamefont{Rogers}},
  \bibinfo{author}{\bibfnamefont{J.}~\bibnamefont{Griffiths}},
  \bibinfo{author}{\bibfnamefont{G.}~\bibnamefont{Jones}},
  \bibinfo{author}{\bibfnamefont{I.}~\bibnamefont{Farrer}},
  \bibinfo{author}{\bibfnamefont{D.}~\bibnamefont{Ritchie}},
  \bibinfo{author}{\bibfnamefont{C.}~\bibnamefont{Smith}},
  \bibinfo{author}{\bibfnamefont{A.}~\bibnamefont{Cottet}},
  \bibnamefont{et~al.}, \bibinfo{journal}{Phys. Rev. Applied}
  \textbf{\bibinfo{volume}{5}}, \bibinfo{pages}{034011} (\bibinfo{year}{2016}).

\bibitem[{\citenamefont{Baart et~al.}(2016)\citenamefont{Baart, Jovanovic,
  Reichl, Wegscheider, and Vandersypen}}]{baart2016nanosecond}
\bibinfo{author}{\bibfnamefont{T.~A.} \bibnamefont{Baart}},
  \bibinfo{author}{\bibfnamefont{N.}~\bibnamefont{Jovanovic}},
  \bibinfo{author}{\bibfnamefont{C.}~\bibnamefont{Reichl}},
  \bibinfo{author}{\bibfnamefont{W.}~\bibnamefont{Wegscheider}},
  \bibnamefont{and} \bibinfo{author}{\bibfnamefont{L.~M.~K.}
  \bibnamefont{Vandersypen}}, \bibinfo{journal}{Appl. Phys. Lett.}
  \textbf{\bibinfo{volume}{109}} (\bibinfo{year}{2016}).

\bibitem[{\citenamefont{Petersson et~al.}(2009)\citenamefont{Petersson, Smith,
  Anderson, Atkinson, Jones, and Ritchie}}]{petersson2009microwave}
\bibinfo{author}{\bibfnamefont{K.}~\bibnamefont{Petersson}},
  \bibinfo{author}{\bibfnamefont{C.}~\bibnamefont{Smith}},
  \bibinfo{author}{\bibfnamefont{D.}~\bibnamefont{Anderson}},
  \bibinfo{author}{\bibfnamefont{P.}~\bibnamefont{Atkinson}},
  \bibinfo{author}{\bibfnamefont{G.}~\bibnamefont{Jones}}, \bibnamefont{and}
  \bibinfo{author}{\bibfnamefont{D.}~\bibnamefont{Ritchie}},
  \bibinfo{journal}{Phys. Rev. Lett.} \textbf{\bibinfo{volume}{103}},
  \bibinfo{pages}{016805} (\bibinfo{year}{2009}).

\end{thebibliography}

\end{document}